\documentclass[11pt,letterpaper]{article}
\pdfoutput=1
\usepackage{jcappub}
\usepackage{bbm,bm}
\usepackage{mathrsfs}
\usepackage{slashed}
\usepackage{caption}
\usepackage{epstopdf}
\usepackage[normalem]{ulem}
\usepackage[bottom]{footmisc}
\usepackage{subcaption}
\usepackage{bbold}
\usepackage{titlesec}
\usepackage{threeparttable}
\usepackage{booktabs}
\usepackage{changepage}
\usepackage[utf8]{inputenc}

\usepackage{grffile}

\usepackage{graphicx}  
\usepackage{dcolumn}   
\usepackage{bm}        
\usepackage{amssymb}   
\usepackage{setspace}
\usepackage{amsmath, amssymb, setspace}
\usepackage{array}
\usepackage{booktabs}
\usepackage{caption}
\usepackage{indentfirst}
\usepackage{float}
\usepackage{lmodern}
\usepackage{multirow}
\usepackage{soul}
\usepackage[normalem]{ulem}

\usepackage{braket}
\usepackage{comment}

\usepackage[draft]{pgf}

\newcommand{\DoBox}[1]{\begin{center}
\color{red}\fbox{
\begin{minipage}{0.9\textwidth}

\end{minipage}}
\end{center}}

\newlength{\myimageoversize}
\newsavebox{\myimage}

\titleformat{\subsubsection}
 {\normalfont\fontsize{12}{17}\itshape}{\thesubsubsection}{1em}{}

\title{\huge{Gravitational wave signatures from discrete flavor symmetries}}

\author[a]{Graciela B. Gelmini,}
\author[b,c]{Silvia Pascoli,}
\author[a]{Edoardo Vitagliano}
\author[d]{and Ye-Ling Zhou}

\affiliation[a]{Department of Physics and Astronomy, University of California, Los Angeles\\
475 Portola Plaza, Los Angeles, CA 90095-1547, USA}
 \affiliation[b]{Institute for Particle Physics Phenomenology, Department of Physics, Durham University, \\Lower Mountjoy, South Rd, Durham DH1 3LE, United Kingdom
}
\affiliation[c]{Dipartimento di Fisica e Astronomia, Universit\`a di Bologna,\\ Via Irnerio 46, I-40126 Bologna, Italy
}

 \affiliation[d]{Department of Physics and Astronomy, University of Southampton\\ University Rd, Southampton SO17 1BJ, United Kingdom
}

\emailAdd{gelmini@physics.ucla.edu}
\emailAdd{silvia.pascoli@durham.ac.uk}
\emailAdd{edoardo@physics.ucla.edu}
\emailAdd{ye-ling.zhou@soton.ac.uk}

\begin{document}

\abstract{
Non-Abelian discrete symmetries have been widely used to explain the patterns of lepton masses and flavor mixing. In these models, a given symmetry is assumed at a high scale and then is spontaneously broken by scalars (the flavons), which acquire vacuum expectation values. Typically, the resulting leading order predictions for the oscillation parameters require corrections in order to comply with neutrino oscillation data. We introduce such corrections through an explicit small breaking of the symmetry. 

This has the advantage of solving the cosmological problems of these models without resorting to inflation. The explicit breaking induces an energy difference or ``bias'' between different vacua and drives the evolution of the domain walls, unavoidably produced after the symmetry breaking, towards their annihilation. Importantly, the wall annihilation leads to gravitational waves which may be observed in current and/or future experiments.
We show that a distinctive pattern  of gravitational waves with multiple overlapped peaks is generated when walls annihilate, which is within the reach of future detectors. We also show that cosmic walls from discrete flavor symmetries can be cosmologically safe for any spontaneous breaking scale between 
1 and  $10^{18}$ GeV, if the bias is chosen adequately, without the need to inflate the walls away. We use as an example a particular $A_4$ model in which an explicit breaking is included in right-handed neutrino mass terms.
}

\maketitle

\section{Introduction}

The discovery of neutrino masses and mixing  \cite{Fukuda:1998mi, Fukuda:2001nj, Ahmad:2001an, Ahmad:2002jz} represents the first laboratory evidence of particle physics beyond the Standard Model (SM). Impressive progress has been made in the past 20 years in determining neutrino oscillation parameters. Two mass squared differences have been measured with best fit values $\Delta m^2_{21} \simeq 7.4\times 10^{-5}~{\rm eV}^2$ and $|\Delta m^2_{31} | \simeq 2.5\times 10^{-3}~{\rm eV}^2$, as well as three mixing angles with $\theta_{12}\simeq 33.4^\circ$, $\theta_{23}\simeq 49.0^\circ$ and $\theta_{13}\simeq 8.6^\circ$\cite{Esteban:2020cvm}. The sign of $\Delta m^2_{31}$ is still unknown, with positive sign corresponding to normal ordering (NO), $m_1< m_2<m_3$, and negative sign to inverted ordering (IO), $m_3< m_1<m_2$. There are some indications in favour of NO, although very recent results have decreased their significance. First hints of leptonic CP-violation due to the $\delta$ phase have been reported~\cite{Esteban:2020cvm}. 

We now know that neutrinos have tiny masses, with at most a mild hierarchy, and that the mixing angles are very different from those in the quark sector. Moreover, we still have to establish if neutrinos are Dirac or Majorana particles. 
These consideration pose fundamental questions in particle physics, in particular concerning the origin of neutrino masses and of leptonic mixing.
Here, we focus on the latter and in particular on the most studied approach to explain the observed values of the leptonic mixing angles, that of non-Abelian discrete symmetries. Many different groups have been considered, for example
$A_4$, $S_4$, $A_5$, just to name a few among the dozen used, with $A_4$ being the most studied example \cite{Ma:2001dn}, and hundreds of realization have been proposed.
This framework assumes that a non-Abelian discrete symmetry unifies flavors at a high energy scale, and its spontaneous symmetry breaking (SSB) gives rise to the flavor mixing. The breaking is achieved by a new type of scalars, called flavons, which gain non-trivial vacuum expectation values (VEV).
Generically, the breaking leaves different conserved subgroups in the charged lepton and neutrino sectors, so that a non-trivial mixing matrix arises in the charged current Lagrangian. Typically, the symmetry induces a leading order mixing pattern, e.g. tribimaximal or bimaximal mixing \cite{Altarelli:2005yp,Altarelli:2005yx}, which is further corrected by small terms in order to achieve a better fit to the data. The most studied one, and closest to data, is the tri-bimaximal (TBM) mixing pattern \cite{Harrison:2002er,Harrison:2002kp,Xing:2002sw} with $\theta_{12}\simeq 35.3^\circ$ and $\theta_{23}=45^\circ$. These values are consistent with current oscillation data at the $3\sigma$ level~\cite{Esteban:2020cvm} 
although the third TBM prediction, $\theta_{13} =0$, was ruled out by the observation of a relatively large
$\theta_{13}$~\cite{An:2012eh, Ahn:2012nd}. In order to explain also this value, various possibilities are considered by including special higher dimensional operators in the direct/semi-direct approach~\cite{King:2013eh}.  A lot of new developments have also been proposed by imposing generalised CP-symmetries \cite{Feruglio:2012cw,Holthausen:2012dk}
and modular invariance~\cite{Feruglio:2017spp}. For the most recent progress of flavor symmetry studies see, e.g.,~\cite{King:2017guk,Xing:2019vks,Feruglio:2019ktm}. 

A critical question is how to test this kind of approach.
At low energy, one can rely on the information provided in the leptonic mixing matrix. Due to the constraints posed by the symmetry, these models typically imply relations between the mixing parameters, known as ``sum-rules"; for a review see e.g. \cite{King:2013eh,King:2014nza}. With sufficiently accurate knowledge of the mixing angles and in particular of the $\delta$ phase, it is possible to test these relations and exclude large classes of models \cite{Ballett:2013wya,Ballett:2014dua,Girardi:2014faa}. This information is very important but somehow indirect, as it does not allow to test directly the mechanism. A direct test would be the observation of the flavons involved in the spontaneous symmetry breaking or some other imprint of it left in the Universe.
As it is assumed that the breaking of the symmetry happens at high scales, this is extremely challenging. If the spontaneous-breaking scale is not much higher than the electroweak one, signatures could arise in the charged lepton sector and in Higgs physics.
Charged-lepton flavor-violating (CLFV) processes such as $\tau \to 3\mu$ and $\mu\to e\gamma$ could 
arise in the framework of flavor symmetries~\cite{Feruglio:2008ht,Ma:2010gs} (see also~\cite{Kobayashi:2015gwa, Muramatsu:2016bda}). 
Assuming these processes are mediated by dimension-six operators with flavor-breaking effects described by a 
flavon VEV, null results of CLFV searches
constrain the flavon symmetry breaking scale up to $\mathcal{O}(10)$ TeV \cite{Feruglio:2008ht}.
In $A_4$ models CLFV processes can also be directly triggered by couplings which are essential to generate lepton  structures~\cite{Pascoli:2016wlt}. Since the couplings are suppressed by lepton masses, the constraints are relaxed. In particular, lower bounds of the flavon masses and spontaneous symmetry breaking  scale in the charged lepton sector are constrained to be around hundreds of GeV. 

Direct searches of flavons at the LHC have also been considered as complementary to constrain the parameter space of $A_4$ models~\cite{Heinrich:2018nip}. Constraints on the scale of the flavor symmetry breaking and flavon masses by adding electroweak interactions for flavons (i.e., arranging flavons as electroweak doublets)  have also been discussed~\cite{Toorop:2010ex, Toorop:2010kt, Holthausen:2012wz, Heeck:2014qea, Varzielas:2015joa}.
However, the vast majority of flavor models in the literature assumes that the symmetry is broken at ultrahigh energy scales and do not consider any possible signature arising from it, apart from the leptonic mixing.
Ultrahigh energy non-Abelian discrete symmetries are also motivated by string theory \cite{Kobayashi:2006wq}, where they arise as a subgroup of the modular group \cite{Altarelli:2005yx,deAdelhartToorop:2011re}
and/or from the orbifolding of extra dimensions \cite{Altarelli:2006kg}. 

Whatever the scale of the discrete flavor symmetry spontaneous breaking, it gives rise  degenerate vacua 
separated by energy barriers
leading to a network of cosmic domain walls. This is a serious shortcoming of these models, as this prediction is in conflict with cosmology if the walls are stable~\cite{Zeldovich:1974uw, Kibble:1976sj, Vilenkin:1984ib}.

Solutions to the domain wall problem
have been discussed in the context of non-Abelian discrete symmetries such as $A_4$. Typical ways are (1) to assume an inflationary era after the spontaneous breaking to inflate  domain walls away, or (2) to  include explicit breaking terms of the discrete symmetry, 
such that the domain walls collapse in  a certain period of time after the spontaneous symmetry breaking~\cite{Riva:2010jm, Antusch:2013toa}. 

Here, we point out the existence of a new potential signature of the discrete flavor symmetry spontaneous breaking, namely gravitational waves (GW) sourced by the collapse of
domain walls in the early Universe~\cite{Vilenkin:1981zs, Preskill:1991kd, Gleiser:1998na, Hiramatsu:2010yz, Kawasaki:2011vv, Hiramatsu:2013qaa}. We will show that GW observations could
 reveal the existence of spontaneously broken  non-Abelian discrete flavor symmetries, for the spontaneous breaking happening from the TeV to the grand unified theory (GUT) scale.
 As a concrete example, we construct an $A_4$ model  where the explicit  $A_4$-breaking  is introduced in the right-handed neutrino mass term. In this model, the explicit breaking term not only is the main source for a non-zero $\theta_{13}$ and CP violation, but also splits the degeneracy of the multiple vacua  that appear after the $A_4$ spontaneous breaking.

\section{Flavor symmetries and mixing: the tetrahedral group $A_4$ as paradigm}\label{sec:vacua}

Non-Abelian discrete symmetries have been widely used to explain the lepton flavor mixing pattern \cite{Ma:2001dn, Altarelli:2005yp,Altarelli:2005yx}.  In this  framework, a non-Abelian discrete group is introduced which acts non-trivially on the flavor space of neutrinos, charged leptons and new scalars, called flavons.  The theory is assumed to be invariant under this group.
 At a certain energy scale, flavons gain vacuum expectation values, and the symmetry is spontaneously broken. 
Different residual symmetries may be preserved approximately in the neutrinos and charged lepton sectors as the flavon responsible for the neutrino mass texture gains a different VEV than the one for the charged lepton mass matrix. The misalignment of these vacua leads to the leptonic flavor mixing in the charged current Lagrangian.
Based on the symmetry argument, it is possible to obtain generic predictions for the leading texture of the mixing without going into the details of the model,
e.g. the TBM mixing in $A_4$ \cite{Altarelli:2005yp,Altarelli:2005yx} and $S_4$ \cite{Lam:2008rs}, bimaximal mixing in $S_4$ \cite{Altarelli:2009gn} and the golden ratio mixing in $A_5$ \cite{Everett:2008et}. 
This is a common feature for model building in the semi-direct and direct approach \cite{King:2013eh}. 
As these mixing patterns are not consistent with current neutrino oscillation data any more, in particular with nonzero $\theta_{13}$, small corrections have to be included. Different types of corrections have been suggested, see \cite{King:2017guk,Xing:2019vks,Feruglio:2019ktm} for some recent reviews.

We will use the tetrahedral group $A_4$ as a paradigm of flavor symmetry. 
This group was first proposed in~\cite{Ma:2001dn} and subsequently extensively studied, e.g., in \cite{Babu:2002dz,Altarelli:2005yx,Morisi:2007ft,Altarelli:2008bg,Feruglio:2008ht,Altarelli:2009kr,Toorop:2010ex,Holthausen:2012wz,Antusch:2013toa,Varzielas:2015joa,King:2018fke,Heinrich:2018nip}. It is the simplest group that has a three-dimensional irreducible representation, allowing the three copies of left-handed lepton doublets to transform universally as a triplet in the flavor space. 
In Subsection~\ref{sec:SSB},  we will discuss the alignment and degeneracy of vacua from the $A_4$  spontaneous symmetry breaking.
Subsection~\ref{sec:mixing} is devoted to a brief review of $A_4$ models in the literature, including the realization of the TBM mixing at leading order and  the necessary corrections at subleading order. 
In Subsection~\ref{sec:explicit} we introduce the explicit breaking, needed to generate sub-leading corrections. We will show that while the $A_4$ SSB is responsible for the large mixing angles $\theta_{12}$ and $\theta_{23}$, the explicit breaking of $A_4$ can produce the correct value of $\theta_{13}$ and consequently generate the CP-violating phase $\delta$.  Readers who are not interested in details of flavor model constructions can skip the rest of this section and go directly to the next.

\subsection{Spontaneous symmetry breaking of $A_4$ \label{sec:SSB}}

$A_4$ is the finite group of even permutations of four objects. It has 12 elements with two generators which satisfy $S^2 = T^3 =
(ST)^3 = \mathbf{1}$. All twelve elements could be represented by $S$ and $T$ as $T$, $ST$, $TS$, $STS$, $T^2$, $ST^2$, $T^2S$, $TST$, $S$, $T^2ST$, $T S T^2$ and the identity element $\mathbf{1}$. The generators $S$ and $T$ in the triplet representation basis can be expressed as 
\begin{eqnarray}\label{eq:MR}
S=\begin{pmatrix}
 1 & 0 & 0 \\
 0 & -1 & 0 \\
 0 & 0 & -1 \\
\end{pmatrix} \,, \quad
T=
\begin{pmatrix}
 0 & 0 & 1 \\
 1 & 0 & 0 \\
 0 & 1 & 0 \\
\end{pmatrix}\,. 
\end{eqnarray}
This group has three $Z_2$ and four $Z_3$ subgroups. The $Z_2$ subgroups are generated by $S$, $TST^2$ and $T^2ST$, respectively.
The $Z_3$ subgroups are generated by $T$, $STS$, $ST$ and $TS$, 
respectively. 
After the $A_4$ SSB, some of these subgroups may be residual in parts of the Lagrangian.
We now discuss the two degeneracy patterns of vacua resulting from the SSB $A_4 \to Z_2$ and $A_4 \to Z_3$, in turn.

The SSB of flavor symmetries are usually achieved by the introduction of new scalars (flavons) which gain non-trivial vacuum expectation values.
In the following, we will introduce a real triplet flavon $\chi = (\chi_1, \chi_2, \chi_3)$ to achieve the breaking of $A_4$ to a $Z_2$ and another flavon $\varphi = (\varphi_1, \varphi_2, \varphi_3)$ to achieve the breaking of $A_4$ to a $Z_3$. 

The tree-level potential for $\chi$ in $A_4$ takes a simple form \cite{Pascoli:2016eld}, 
\begin{eqnarray} \label{eq:self-coupling1}
V_{\rm tree}(\chi) = \frac{1}{2}\mu_\chi^2 I_1(\chi) + \frac{g_1}{4} I_1^2(\chi)+ \frac{g_2}{4} I_2(\chi)\, , 
\end{eqnarray}
where $\mu^2_\chi$, $g_1$ and $g_2$ are real parameters, and
\begin{eqnarray} \label{eq:I_i}
I_1(\chi) &\equiv& \chi_1^2 + \chi_2^2 + \chi_3^2 \, , \nonumber\\
I_2(\chi) &\equiv& \chi_1^2\chi_2^2+\chi_2^2\chi_3^2+\chi_3^2\chi_1^2 \, .
\end{eqnarray}
Here, the coefficients $\mu_\chi^2<0$ and $g_1>\max\{ 0, g_2/3 \}$ are assumed to ensure the $A_4$ symmetry spontaneously breaks to a non-trivial stable vacuum. Note that there should be an additional cubic term $\chi_1\chi_2\chi_3$ which is $A_4$-invariant. For simplicity, we assume it is small, corresponding to an approximate parity symmetry, $\chi \leftrightarrow -\chi$.

We consider the vacuum configuration of $\chi$ derived
from this potential. All vacua can be found by minimizing the potential.
A complete list of vacua for this type of potential have been obtained and listed in \cite{Pascoli:2016eld}. In order to force $A_4$ to be spontaneously broken to a residual $Z_2$ symmetry, an additional requirement $g_2 > 0$ has to be imposed. The full list of $Z_2$-invariant vacua are given by
\begin{eqnarray} \label{eq:vev_chi}
{\bm v}_{\mathbf{1}}^{\pm}= \pm v_\chi \begin{pmatrix} 1 \\ 0 \\ 0 \end{pmatrix}, ~
{\bm v}_{\mathbf{2}}^{\pm}= \pm v_\chi \begin{pmatrix} 0 \\ 1 \\ 0 \end{pmatrix},~
{\bm v}_{\mathbf{3}}^{\pm}= \pm v_\chi \begin{pmatrix} 0 \\ 0 \\ 1 \end{pmatrix}  \,,
\end{eqnarray}
where $v_\chi$ now is fixed at $v_\chi= \sqrt{-\mu^2_\chi/g_1}$ and required to be positive. The vacua ${\bm v}_{\mathbf{1}}^\pm$, ${\bm v}_{\mathbf{2}}^\pm$ and ${\bm v}_{\mathbf{3}}^\pm$ preserve different residual $Z_2$ symmetries generated by $S$, $TST^2$ and $T^2ST$, respectively. Namely, ${\bm v}_{\mathbf{1}}^{\pm}$ is invariant under the transformation of $S$, $S{\bm v}_{\mathbf{1}}^{\pm} = {\bm v}_{\mathbf{1}}^{\pm}$. 
These vacua are physically equivalent unless explicit $A_4$-breaking terms are introduced, as we will discuss later.

We further consider the SSB of $A_4 \to Z_3$. 
The tree-level renormalizable self-couplings for $\varphi$ takes a similar form as for $\chi$, 
\begin{eqnarray} \label{eq:self-coupling}
V_{\rm tree}(\varphi) = \frac{1}{2}\mu^2_\varphi I_1(\varphi) + \frac{f_1}{4} I_1^2(\varphi)+ \frac{f_2}{4} I_2(\varphi)\,, 
\end{eqnarray}
where $\mu^2_\varphi$, $f_1$, $f_2$ are real parameters, and $I_1$ and $I_2$ have been defined in Eq.~\eqref{eq:I_i}. Again, the coefficients $\mu^2_\varphi<0$ and $f_1>\max\{ 0, f_2/3 \}$ have to be assumed to ensure the $A_4$ symmetry spontaneously breaks to a non-trivial stable vacuum. In order to achieve the spontaneous breaking of $A_4$ to $Z_3$ instead of $Z_2$, the restriction $f_2 < 0$ is required. The degenerate vacua are given by
\begin{eqnarray} \label{eq:vev_varphi}
{\bm u}_{\mathbf{1}}^{\pm}= \pm \frac{v_\varphi}{\sqrt{3}} \begin{pmatrix} 1 \\ 1 \\ 1 \end{pmatrix}\,, \quad
{\bm u}_{\mathbf{2}}^{\pm}= \pm \frac{v_\varphi}{\sqrt{3}} \begin{pmatrix} -1 \\ 1 \\ 1 \end{pmatrix}\,,\quad
{\bm u}_{\mathbf{3}}^{\pm}= \pm \frac{v_\varphi}{\sqrt{3}} \begin{pmatrix} 1 \\ -1 \\ 1 \end{pmatrix}\,,\quad
{\bm u}_{\mathbf{4}}^{\pm}= \pm \frac{v_\varphi}{\sqrt{3}} \begin{pmatrix} 1 \\ 1 \\ -1 \end{pmatrix}\,,
\end{eqnarray}
where $v_\varphi$ is fixed at $v_\varphi = \sqrt{-\mu_\varphi^2/(f_1+f_2/3)}$. These vacua  preserve $Z_3$ symmetries generated by $T$, $STS$, $ST$ and $TS$, respectively.

\begin{figure}[htb]
\begin{center}
\includegraphics[width=.98\textwidth]{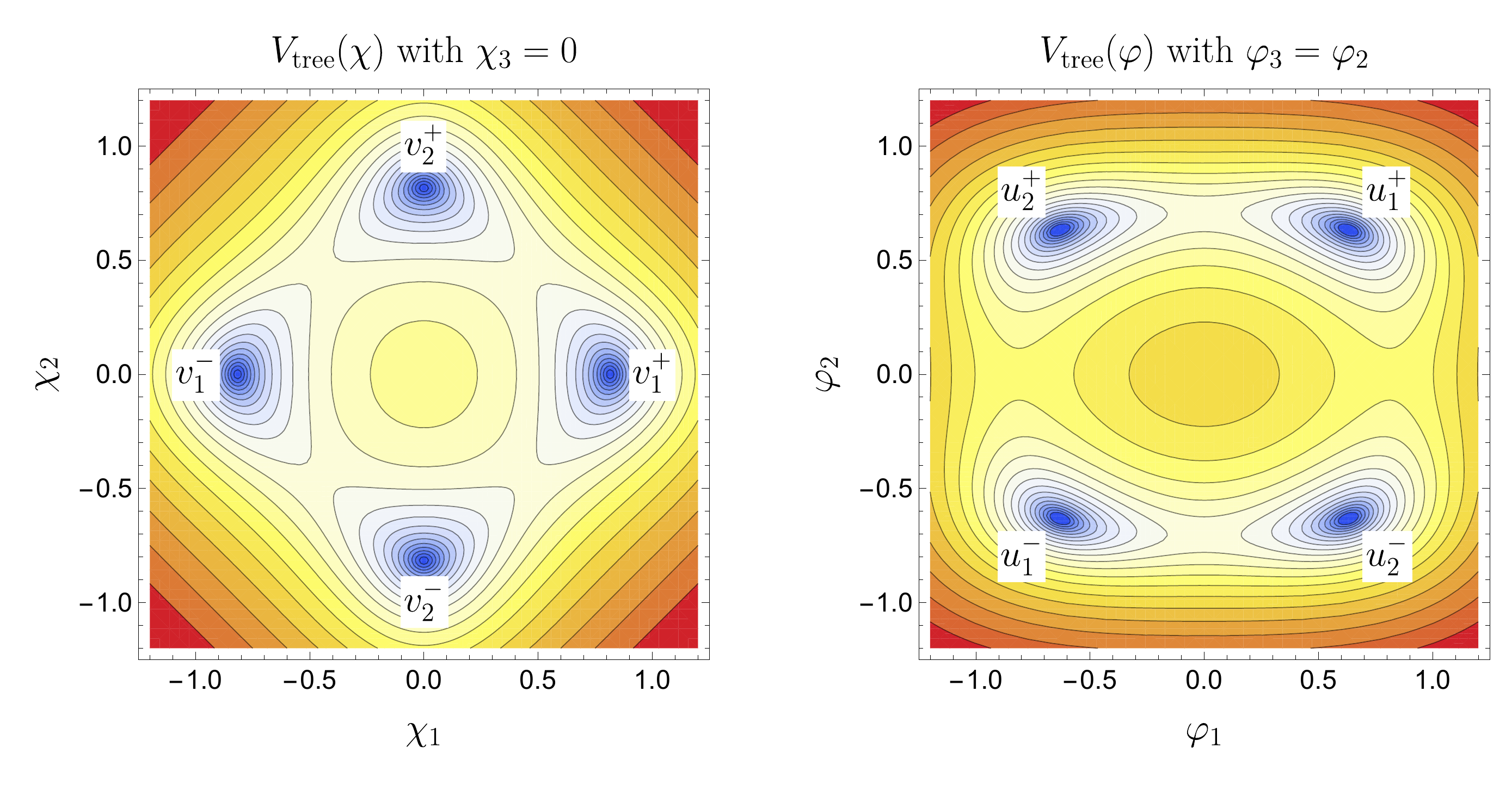}
\caption{ \label{fig:potential} Contour plots of potentials:  (a) $V_{\rm tree}(\chi)$ as a function of $\chi_1$ and $\chi_2$ with $\varphi_3=0$ fixed; (b) $V_{\rm tree}(\varphi)$ with $\varphi_3=\varphi_2$ as a function of $\varphi_1$ and $\varphi_2$ with $\chi_3=\varphi_2$ fixed. Inputs: $\mu_\chi=\mu_\varphi = 1$, $g_1 = f_1 = 3/2$ and $g_2 = 4$, $f_2 = -2$. Degenerate vacua ${\bm v}_{\bm 1}^{\pm}$, ${\bm v}_{\bm 2}^{\pm}$ and ${\bm u}_{\bm 1}^{\pm}$, ${\bm u}_{\bm 2}^{\pm}$ are shown in the plot.}
\end{center}
\end{figure}

The potential and the degeneracy of vacua can be interpreted easily in a geometrical way. In Figure~\ref{fig:potential}, we show contour plots of $V_{\rm tree}(\chi)$ and $V_{\rm tree}(\varphi)$ as function of $(\chi_1, \chi_2)$ and $(\varphi_1, \varphi_2)$, respectively, with $\chi_3=0$, $\varphi_3 = \varphi_2$ fixed. Minima of $V_{\rm tree}(\chi)$ at four vacua ${\bm u}_{\bm 1}^+$, ${\bm u}_{\bm 2}^+$, ${\bm u}_{\bm 1}^-$ and ${\bm u}_{\bm 2}^-$ and those of $V_{\rm tree}(\varphi)$ at ${\bm u}_{\bm 1}^+$, ${\bm u}_{\bm 2}^+$, ${\bm u}_{\bm 1}^-$ and ${\bm u}_{\bm 2}^-$ are indicated in the figure. 
The degenerate vacua are disconnected with each other in the flavon space.

In the above, we have discussed vacuum alignments led by only self couplings of the potential for a single flavon triplet (either $\chi$ or $\varphi$). In a flavor model, as will be shown in Section~\ref{sec:mixing}, several flavon multiplets may be involved together to generate flavor mixing, and some of their cross couplings could not be avoided by the symmetry. One term which cannot be forbidden for the simplified model in Section~\ref{sec:mixing} is 
 $(\varphi_1^2+\omega \varphi_2^2+\omega^2 \varphi_3^2) (\chi_1^2+\omega^2 \chi_2^2+\omega \chi_3^2) +{\rm h.c.}$ This coupling has been discussed in \cite{Pascoli:2016eld}. It can shift $\chi$ and $\varphi$ VEVs and the mixing, and in particular, provide a source for a non-zero $\theta_{13}$. With a suitable choice of the coefficient, this coupling can give a correct prediction for the value $\theta_{13} \simeq 8.6^\circ$. 
Plenty of $A_4$ models have also considered cross couplings from higher-dimensional operators (for recent reviews see, e.g., \cite{King:2014nza,Xing:2019vks,Feruglio:2019ktm}). They may also break the residual symmetries in either the charged lepton or the neutrino sector and further shift the lepton flavor mixing. 
Another type of cross couplings are those between flavons and the SM Higgs field, i.e., $H^\dag H I_1(\chi)$ and $H^\dag H I_1(\varphi)$ cannot be forbidden by $A_4$. These couplings may result in shifts 
of the $v_\chi$ and $v_\varphi$ values after  electroweak SSB, and could be used in collider flavon searches~\cite{Heinrich:2018nip}. Following a widely assignment that the Higgs is  a singlet of the flavor symmetry, 
these couplings do not affect the flavor structure of the
flavon VEV.

In this paper, we simplify our discussion by assuming all cross couplings to be negligibly small. For cross couplings between flavons, we make this assumption due to the phenomenological interest that we will
include the $A_4$ explicit breaking as a new source of $\theta_{13}$. 
Cross couplings between flavons and the Higgs are necessarily assumed to be small to avoid significant modifications to the Higgs sector. 
On the other hand, although some cross couplings cannot be forbidden by the symmetry, all these couplings can be forbidden by extending the geometry to extra dimensions \cite{Altarelli:2005yp} or by  supersymmetry \cite{Altarelli:2005yx}. 

\subsection{Flavor mixing in $A_4$ \label{sec:mixing}}

We now briefly discuss how leptonic TBM mixing and its possible corrections is realized in $A_4$ models. 
The three SM lepton doublets $L=(L_1, L_2, L_3)$ are arranged as a triplet of $A_4$, and right-handed charged leptons $e_R$, $\mu_R$ and $\tau_R$ are taken as singlets $\mathbf{1}$, $\mathbf{1}'$ and $\mathbf{1}''$, respectively. Three copies of right-handed neutrinos $N=(N_1,N_2,N_3)$ 
are also arranged as a triplet of $A_4$. The two real $A_4$-triplet flavons $\chi$ and $\varphi $ discussed in the Section~\ref{sec:vacua} will be used to generate the flavor structures in the neutrino and charged lepton sectors, respectively.  
We take the same $A_4$-invariant Lagrangian terms for lepton mass generations used in~\cite{Pascoli:2016eld}, which is also widely used in this type of models. They can be written as\footnote{Explicit forms of these Lagrangian terms depend on the chosen representation basis, see Appendix~A for details.} 
\begin{eqnarray}
-\mathcal{L}_{l,\nu} &\supset& 
y_D \bar{L}_i \tilde{H} N_i + 
y_N \bar{N}_i N_j^c \chi_k + 
\frac{1}{2}u \bar{N}^c_i N_i \nonumber\\
&+& 
 \frac{\varphi_i}{\Lambda}\bar{L}_i H (y_e e_R + 
\omega^{1-i} y_\mu \mu_R + 
\omega^{i-1} y_\tau \tau_R) +  
\text{h.c.} \,, 
\label{eq:Lagrangian}
\end{eqnarray}
where $i,j,k$ sum for $1,2,3$ with $i \neq j \neq k \neq i$ and $\omega = e^{i 2\pi/3}$, $y_{e,\mu,\tau}$, $y_D$ and $y_N$ are dimensionless coefficients and $u$ is a mass-dimension parameter. Note that additional $Z_n$ symmetries are necessary to forbid unnecessary terms, see \cite{Pascoli:2016eld} for details. To simplify our discussion, we impose a CP-symmetry, which enforces all coefficients $y_{e,\mu,\tau}$, $y_D$, $y_N$ and $u$ to be real. 

Flavons gain VEV as outlined in the previous subsection. After the  $A_4$ SSB, $\chi$ and $\varphi$ have the
VEV in Eqs.~\eqref{eq:vev_chi} and \eqref{eq:vev_varphi}, respectively. 
Here, without loss of generality, we choose 
\begin{eqnarray}
\langle \chi \rangle = {\bm v}_{\bf 1}^+\,, \quad
\langle \varphi \rangle = {\bm u}_{\bf 1}^+ \,.
\end{eqnarray} 
The  $\chi$ VEV lead to the right-handed Majorana neutrino mass matrix of the from
\begin{eqnarray} \label{eq:mass_N} 
M_N = \begin{pmatrix} 
u & 0 & 0 \\
0 & u & y_N v_\chi \\
0 & y_N v_\chi & u
\end{pmatrix}\,. 
\end{eqnarray} 
In the charged lepton sector, the Yukawa coupling $\bar{L} Y_l H (e,\mu,\tau)_R^T$ is generated after $\varphi$ gains the VEV, where the $3\times 3$ Yukawa coupling matrix is given by 
\begin{eqnarray} \label{eq:Yukawa_l} 
Y_l = \frac{v_\varphi}{\sqrt{3}\Lambda}\begin{pmatrix} 
y_e & y_\mu & y_\tau \\
y_e & \omega^2 y_\mu & \omega y_\tau \\
y_e & \omega y_\mu & \omega^2 y_\tau
\end{pmatrix}
\end{eqnarray}
with $\omega = e^{i 2\pi/3}$. After the SM Higgs gains the VEV $\langle H \rangle = v_H = 174$~GeV, the mass matrix for charged leptons $M_l = Y_l v_H$, and the light neutrino matrix $M_\nu = y_D^2 v_H^2 M_N^{-1}$ arise. 
The charged lepton mass matrix is diagonalised via $U_l^\dag M_l M_l^\dag U_l = {\rm diag} \{ m_e, m_\mu, m_\tau \}$, and the light neutrino mass matrix via $U_\nu^\dag M_\nu U_\nu^* = {\rm diag} \{ m_1, m_2, m_3 \}$, with  unitary matrices $U_l$ and $U_\nu$. Their product $U \equiv U_l^\dag U_\nu$ enters in the SM charged current lagrangian and is denoted as the lepton flavor mixing matrix, whose values can be measured in neutrino oscillation experiments.
With the right-handed neutrino mass matrix given in Eq.~\eqref{eq:mass_N} and the charged lepton Yukawa terms given in Eq.~\eqref{eq:Yukawa_l}, we find the mixing to be tri-bimaximal 
\begin{eqnarray} \label{eq:TBM}
|U| = \left(
\begin{array}{ccc}
 \frac{2}{\sqrt{6}} & \frac{1}{\sqrt{3}} & 0 \\
 \frac{1}{\sqrt{6}}& \frac{1}{\sqrt{3}} & \frac{1}{\sqrt{2}} \\
 \frac{1}{\sqrt{6}}& \frac{1}{\sqrt{3}} & \frac{1}{\sqrt{2}} \\
\end{array}
\right) \,.
\end{eqnarray} 
The TBM mixing is well-known for its prediction of mixing angles $\theta_{12}\simeq 35.3^\circ$ and $\theta_{23}=45^\circ$~\cite{Harrison:2002er,Harrison:2002kp,Xing:2002sw}, which are still consistent with neutrino oscillation data.
The prediction of TBM in $A_4$ models is not accidental but mostly guaranteed by the symmetry. In detail, it is a consequence of a residual $Z_3$ symmetry preserved in the charged lepton sector, and a $Z_2\times Z_2^{\mu\tau}$ in the neutrino sector \cite{Lam:2008rs}. Here, $Z_2^{\mu\tau}$ refers to the $\nu_\mu$-$\nu_\tau$ permutation symmetry. It is not a sub-symmetry of $A_4$, but is preserved in the most general Lagrangian compatible with $A_4$ and the specified representation content for the flavons \cite{Altarelli:2010gt}. However, as the  vanishing $\theta_{13}$ predicted by TBM has been ruled out by neutrino oscillation data, corrections have to be introduced. 

Mass eigenvalues for charged leptons and right-handed neutrinos are found to be $m_{e,\mu,\tau} = |y_{e,\mu,\tau}|(v_\varphi v_H/\Lambda)$, $M_{1,3} = |u \pm y_N v_\chi|$ and $M_2 = |u|$. 
Light neutrino mass eigenvalues are simply given by $m_i = y_D^2 v_H^2 M_i^{-1}$ for $i=1,2,3$. 
Neutrino masses $m_1$, $m_2$ and $m_3$ are correlated with each other by $u$, $y_N v_\chi$, and their relative phase. The mass correlation was first predicted in \cite{Altarelli:2005yx}, and has been achieved in dozens of models (see reviews \cite{Barry:2010yk, Gehrlein:2016wlc} and references therein). By tuning these parameters, the model can predict mass spectra consistent with data. 
Instead of a complete scan of the parameter space,
we show some typical values of neutrino masses for  all coefficients real \cite{Feruglio:2012cw,Holthausen:2012dk}. 
We obtain the following sum rules for the neutrino masses,
\begin{eqnarray} \label{eq:sum_rule}
&& \frac{2}{m_2} = \frac{1}{m_1} + \frac{1}{m_3},\quad 
\frac{2}{m_2} = \frac{1}{m_1} - \frac{1}{m_3},\quad 
\frac{2}{m_2} = - \frac{1}{m_1} + \frac{1}{m_3}\, ,\nonumber\\
\nonumber\\
&& 
2M_2 = M_1 + M_3,\quad 
2M_2 = M_1 - M_3,\quad
2M_2 = - M_1 + M_3\,.
\end{eqnarray}
Using the current best fits of oscillation data for both NO and IO light neutrino mass ordering, we obtain three solutions, denoted below as NO1, NO2 and IO1,
\begin{eqnarray} \label{eq:mass_ratio}
&&(m_1, m_2, m_3) = \left\{ \begin{array}{ll}
(0.0058, 0.0104, 0.0506)~{\rm eV}\,, & \quad \mbox{for NO1},  \\
(0.0044, 0.0097, 0.0504)~{\rm eV}\,, & \quad \mbox{for NO2},  \\
(0.0532, 0.0539, 0.0179)~{\rm eV}\,, & \quad \mbox{for IO1} .
\end{array} 
\right.
\end{eqnarray}
The ratios of $M_1$, $M_2$ and $M_3$ in these cases are given respectively by
\begin{eqnarray} \label{eq:mass_ratio2}
&&M_1: M_2: M_3 = \left\{ \begin{array}{ll}
1: 0.558: 0.114\,, & \quad \mbox{for NO1}, \\
1: 0.456: 0.087\,, & \quad \mbox{for NO2}, \\
1: 0.987: 2.974\,, & \quad \mbox{for IO1}.
 \end{array} \right.
\end{eqnarray} 
In the NO case, the mass of the heaviest right-handed neutrino $N_1$ is one order of magnitude higher than that of the lightest right-handed neutrino $N_3$. In the IO case, the mass of the heaviest one $N_3$ is only three times larger than the lightest one $N_2$. Independently of the mass ordering, we denote the heaviest right-handed neutrino mass as $M_N$. The Yukawa coupling between right-handed neutrinos and the flavon is determined to be $y_N = M_N/v_\chi \times 0.443 (0.543) [0.668]$ for NO1 (NO2) [IO1]. 

One may check that choosing a different sets of VEV of $\chi$ and $\varphi$ may lead to different flavor structures for $M_N$ and $Y_l$, and consequently for $M_\nu$ and $M_l$ in the given representation basis. However, the prediction for $|U|$ is the same.

The flavor model constructed here is not UV complete but includes the basic features of most $A_4$ models, namely the realization of the TBM mixing at leading order. We have simplified it in several ways.
To begin with, we ignored additional Abelian symmetries and associated new particles. These new degrees of freedoms could be necessary to forbid unnecessary terms, see~\cite{Pascoli:2016eld} for details. 
Another simplification is that terms for charged lepton mass generation have been written in the form of higher dimensional operators. This form has been widely used in $A_4$ models. These terms can be renormalized by introducing heavy mediators. A simple way is introducing singly-charged vector-like leptons with representation arrangement in the flavor space the same as that of $L$ with the Lagrangian terms
\begin{eqnarray}
-\mathcal{L}_l &\supset& y_E  \bar{L}_i H E_{iR} + M_E \bar{E}_{iL} E_{iR} + \bar{E}_{iL} \varphi_i (y_e e_R + 
\omega^{1-i} y_\mu \mu_R + 
\omega^{i-1} y_\tau \tau_R)\,.
\end{eqnarray}
Lastly, the cross coupling between flavons and the Higgs field $H^\dag H I_1(\chi)$ and $H^\dag H I_1(\varphi)$ cannot be forbidden by $A_4$, however we will avoid addressing
the gauge hierarchy problem. Therefore, as a simplification,  we assume $g_{H\chi} \ll v_H^2/v_\chi^2$ so that the tree-level correction to the Higgs mass is $\delta m_h^2 \ll m_H^2$. This coupling may result in a small shift 
of the $v_\chi$ value after the electroweak breaking, and could be used in collider flavon searches~\cite{Heinrich:2018nip}. However, since the Higgs is a  singlet of the flavor symmetry,
it does not contribute to any flavor structure of the
$\chi$ VEV.

\subsection{$\theta_{13}$ and CP violation arising from  explicit breaking \label{sec:explicit}}

We consider here how to generate a non-zero $\theta_{13}$ from the explicit $A_4$ breaking. 
We include an explicit breaking in the right-handed neutrino mass term
\begin{eqnarray} \label{eq:A4_breaking_source}
-\mathcal{L}_{\slashed{A}_4} &=&  
\frac{1}{2} \epsilon_{ij} v_\chi \bar{N}^c_i N_j  +  
\text{h.c.}  
\label{eq:Lagrangian_breaking}
\end{eqnarray}
Here, the dimensionless parameters $\epsilon_{ij}$ represent the relative size of the explicit breaking. They are assumed to be small. Therefore, $A_4$ can be regarded as a good approximate symmetry and satisfies ’t Hooft’s naturalness criterion \cite{tHooft:1979rat} as switching off these small parameters recovers the  symmetry. There is no further guiding principle to choose their values. 
As an economical choice to agree with neutrino oscillation data and have a unique true vacuum, we assume that
$\epsilon_{22}$, $\epsilon_{33}$ and $\epsilon_{23}$ do not vanish, and $-\epsilon_{22}=\epsilon_{33}$, which leads to ${\bm v}^+_{\bm 1}$ as the true vacuum.
Including these terms, the right-handed neutrino mass matrix  is modified to 
\begin{eqnarray} \label{eq:mass_matrices}
&&M_N = \begin{pmatrix} 
u~~~ & 0~~~ & 0 \\
0~~~ & u + \epsilon_{22} v_\chi~~~ & y_N v_\chi + \epsilon_{23} v_\chi \\
0~~~ & y_N v_\chi + \epsilon_{23} v_\chi~~~ & u - \epsilon_{22} v_\chi
\end{pmatrix}\,. 
\end{eqnarray} 
One can confirm that although the mass eigenvalues for right-handed neutrinos and those for light neutrinos are corrected, 
the sum rule in Eq.~\eqref{eq:sum_rule} is still satisfied explicitly. As a consequence, the mass ratios in Eqs.~\eqref{eq:mass_ratio} and \eqref{eq:mass_ratio2} remain valid. 
The mixing matrix is corrected, taking the form
\begin{eqnarray} \label{eq:TM2}
|U| = \left(
\begin{array}{ccc}
 \frac{2}{\sqrt{6}} \cos \theta~~~ & \frac{1}{\sqrt{3}}~~~ & \sqrt{\frac{2}{3}} |\sin \theta|\\
 |\frac{1}{\sqrt{6}}\cos \theta - \frac{i}{\sqrt{2}} \sin \theta|~~~ & \frac{1}{\sqrt{3}}~~~ & |\frac{1}{\sqrt{6}}\sin \theta + \frac{i}{\sqrt{2}} \cos \theta| \\
 |\frac{1}{\sqrt{6}}\cos \theta - \frac{i}{\sqrt{2}} \sin \theta|~~~ & \frac{1}{\sqrt{3}}~~~ & |\frac{1}{\sqrt{6}}\sin \theta - \frac{i}{\sqrt{2}} \cos \theta| \\
\end{array}
\right) \,,
\end{eqnarray} 
where $\theta$ is a real angle used to diagonalize $M_N$. This model predicts the well-known TM$_2$ mixing, in which $\theta_{12}$ and $\theta_{13}$ are correlated with the sum rule $\sin\theta_{12} \cos\theta_{13} = 1/\sqrt{3}$ \cite{Albright:2010ap}.
$\theta_{13}$ is approximately given by
\begin{eqnarray} \label{eq:theta_13}
\sin\theta_{13} 
\simeq 2 \sqrt{\frac{2}{3}} \frac{ |\epsilon_{22}| v_\chi M_2}{|\Delta M_{31}^2|} \, ,
\end{eqnarray}
where $\Delta M_{31}^2 = M_3^2-M_1^2$.
In the limit $\epsilon_{22} \to 0$, we recover  vanishing $\theta_{13}$. 
Furthermore, it also complies with the well-known $\mu$-$\tau$ reflection symmetry \cite{Harrison:2002et}, where $\theta_{23}$ and all CP-violating phases take the simple values $\theta_{23} = 45^\circ$, $\delta = - 90^\circ$,\footnote{Another prediction of the $\mu$-$\tau$ reflection symmetry is $\delta=+ 90^\circ$, but it is not consistent with oscillation data at $3\sigma$ confidence level, so we do not consider it any further.
} $\alpha_{21}, \alpha_{31} = 0 ~{\rm or}~ 180^\circ$~\cite{Zhou:2014sya}. This mixing pattern has been well studied (for a review see, e.g.~\cite{Xing:2015fdg}), and thus we will not discuss it any further.
Note that the realization of TM$_2$ and $\mu$-$\tau$ reflection mixing depends on
the particular choice of the explicit breaking. 
The main point is that while the TBM is usually generated from the spontaneous breaking of $A_4$, subleading corrections, in particular, the non-zero $\theta_{13}$ and CP violation, can be generated from a small explicit breaking source in the right-handed neutrino masses.

\section{Evolution of cosmic walls}

The spontaneous breaking of a discrete symmetry leads to the formation of topological defects separating different
degenerate ground states of the potential, the cosmic domain walls. Since the initial 1974 work of Zel'dovich, Kobzarev, and Okun~\cite{Zeldovich:1974uw}, it was understood that the formation of walls in the early Universe is  cosmologically unacceptable, unless they have disappeared early enough. The problem is that the energy density of cosmic walls could dominate the total energy density of the Universe, producing a power law inflation. Thus, walls need to disappear before their energy would become dominant. Also, even if walls are subdominant, their presence in the present Universe would cause distortions in the cosmic microwave background (CMB) which conflicts with the present observations, unless  the spontaneous symmetry breaking scale of the walls is below an MeV~\cite{Lazanu:2015fua}, as first pointed out in~\cite{Zeldovich:1974uw}. The possibility that walls annihilate early enough due to a small energy difference between the minima of the potential was also first suggested by Zel'dovich et al.~\cite{Zeldovich:1974uw}  and further studied shortly after (see e.g.~\cite{Zeldovich:1974uw, Kibble:1976sj, Vilenkin:1981zs, Sikivie:1982qv, Gelmini:1988sf, Preskill:1991kd,Chen:2020wvu}).  If the discrete symmetry is not exact, there is an energy difference between the vacua at the two sides of each wall, hereafter a ``bias'', that  causes the false-vacuum regions to disappear.

General equations to provide order of magnitude estimates of the breaking and bias  energy scales are better derived from a simple model. Consider  the Lagrangian of a real scalar field $\phi$ with a $Z_2$ symmetry,
\begin{equation}
 \label{toy-model}
\mathcal{L} = -\frac{1}{2}\partial^{\mu}\phi\partial_{\mu}\phi -  \frac{\lambda}{4}\left(\phi^2-v^2\right)^2 \, ,
\end{equation}
whose potential has two minima at $\langle\phi\rangle=\pm v$ ($v$ is the magnitude of the $\phi$ VEV).
The height of the barrier between the minima is $\lambda v^4/4$. When the discrete  $Z_2$ symmetry is broken, regions of the Universe with different minima  are separated by a domain wall.  The width of the wall results from a  balance between the potential energy (which tends to make the wall thinner) and the gradient term (which tends to make it wider) and in this simple case is (see e.g.~\cite{Gelmini:1988sf})
\begin{equation}
 \label{wall-thickness}
\Delta= (\sqrt{\lambda/2}~v)^{-1} \, .
\end{equation}
Integrating the 00 component of the wall  stress-energy tensor (see e.g.~\cite{Gelmini:1988sf}), one derives the surface tension, which is equal to the energy per unit area of the wall in its rest frame,
\begin{equation}
 \label{sigma}
\sigma = \frac{2 \sqrt{2}}{3} \lambda^{1/2} \, v^3 \equiv f_\sigma \, v^3 \, .
\end{equation}
In the second equality we define the dimensionless real positive constant $f_\sigma$, which characterizes the surface tension. This allows us to write $\sigma$ in terms of $v$ in any model. While $f_\sigma= \mathcal{O}(1)$ in the model of Eq.~\eqref{toy-model}, it could be smaller in more realistic models.

We are going to assume that a small explicit breaking introduces a bias,
\begin{equation}
 \label{bias}
V_{\rm bias} \equiv \epsilon_b \, v^4 \, 
\end{equation}
between the two vacua. Here, $\epsilon_b$ is a real positive dimensionless constant that characterizes the bias and $\epsilon_b \ll 1$. 

Without significant friction on the walls, which we will consider in Section~\ref{viscosity}, their evolution is entirely determined by the wall tension and the bias.

The zero-temperature potential in Eq.~\eqref{toy-model} is only valid for temperatures safely below the phase transition critical temperature $T_c$.  We assume that the Universe is radiation dominated before the phase transition, as well as after it, during the whole evolution of the walls. Using the temperature-corrected potential it is easy to prove that for our toy model   $T_c= 2 v$~\cite{Gelmini:1988sf, Kibble:1976sj}. We are going to assume that $T_c \simeq v$ in all models we consider.

 At $T \ge T_c$  the minimum of the potential is at $\phi=0$. As $T$ approaches  $T_c$, thermal fluctuations in the field become large and regions fluctuate rapidly between the two lower temperature vacua.  As $T$ decreases below $T_c$, the barrier between the two vacua increases, fluctuations become progressively rarer, and when they become exponentially suppressed, patches of different vacua become fixed. These patches have characteristic linear size $R\simeq \xi$, where the correlation length $\xi$ is approximately given by the  inverse temperature dependent mass of the field for $T< T_c$,    $\xi \simeq  [v\, \sqrt{\lambda (1- T^2/ T_c^2) }~]^{-1}$ for our toy model.  So as $T$ decreases below $T_c$,  the average distance between walls, which is also the average radius of curvature of the walls, becomes rapidly $R \simeq  ( \sqrt{\lambda}\, v)^{-1}$~\cite{Kibble:1976sj}. After the phase transition,  each $\xi^3$-cell has a certain probability of being occupied by a particular vacuum. As the bias increases, the probability of false vacua diminishes. Domains of horizon size of a particular vacuum exist only if the probability of a $\xi^3$-cell to be in that vacuum is above a percolation probability. The limit $\epsilon_b \lesssim \mathcal{O}(0.1 \, \lambda)$~\cite{Gelmini:1988sf} ensures that domains of all vacua percolate, so at the completion of the phase transition there will be at least one domain wall extending to the horizon.
 
 The subsequent motion of the walls is determined by  the surface tension and the energy difference between the vacua. The surface tension tends to rapidly straighten out curved walls to the horizon scale, which is of the order of the lifetime of the Universe  $t_U$, and produces a pressure $p_T \simeq \sigma/ R$. The volume pressure due to the energy difference between the two vacua $p_V \simeq V_{\rm bias} =\epsilon_b v^4$  tends to accelerate the walls towards the false vacuum, converting false vacuum into true vacuum. The energy released in this conversion fuels the wall motion. Notice that $p_T$ coincides with the energy density stored in the walls $\rho_w \simeq \sigma R^2/ R^3$ and $p_V$  with the energy density of the false vacuum (with the true vacuum energy being zero).   
 
Different evolution histories of the walls can happen depending on the relative size of $p_T$ and $p_V$ (see e.g.~\cite{Gelmini:1988sf}).  Assuming that initially $p_V \ll p_T$, before $p_V$ starts acting the wall tension  brings $R$  very fast to the horizon size $R\simeq t_U= (1/2) H^{-1}$. Here, $H = \sqrt{ (8 \pi^3 g_\ast/90)} ~T^2/ M_{Pl}= 1.7 ~g_\ast^{1/2}~ T^2/ M_{Pl}  \simeq g_\ast^{1/2} T^2/ M_{Pl}$ is the Hubble expansion rate for a radiation dominated Universe, whose density is $\rho_{r} = (\pi^2/30) ~ g_\ast T^4$, and  $M_{Pl}= G^{-1/2}$ is the Planck mass. If only  the tension force acted on the walls, right after their formation it would induce oscillations of frequency $R^{-1}$. These would quickly accelerate the walls to relativistic speeds. However, this motion is actually damped by emission of particles and by friction due mostly to flavons reflection on the walls (see below).  So the walls become flatter  to the horizon size and slow down, within a Hubble evolution time.

Thus, in the following we will assume that right after the phase transition $R (T_c) \simeq  (1/2) H(T_c)^{-1}$. 
 For this to happen, $p_V$ must be negligible at this point, i.e.  $p_V <p_T  \simeq 2 \sigma  H(T_c)$, which implies
\begin{equation}
 \label{pVupperlimit}
\epsilon_b  < 1 \times 10^{-15} \left(\dfrac{ f_\sigma\, v}{ \mathrm{TeV}}\right) \left(\dfrac{g_*(T_{c})}{10}\right)^{1/2} \, .
\end{equation}
 After the phase transition, while $p_V<p_T$ the walls enter into a scaling regime,  in which $R (T)$ is always of the size of the horizon (see e.g.~\cite{Vilenkin:2000jqa, Hindmarsh:1996xv,  Hindmarsh:2002bq}), as confirmed by numerical 
 studies~\cite{Press:1989yh,Garagounis:2002kt,Oliveira:2004he,Avelino:2005kn,Leite:2011sc,Leite:2012vn, Martins:2016ois}. In the scaling  regime, the energy density stored in the walls  would become larger than the critical density,  $\rho_w > \rho_{c}=3 H^2/ 8 \pi G= 3 M_{Pl}^2/ 32 \pi t^2 $,  at
\begin{equation}
\label{tdom} 
   t_{\rm{dom}} \simeq \dfrac{M_{Pl}^2}{33 ~\sigma} \,  ,
 \end{equation}
in a radiation dominated Universe, for which $\rho_{c}= \rho_{r}$.

We thus impose that  the volume pressure becomes important before $t_{\rm{dom}}$  and drives the walls to annihilate at $t_{\rm{ann}} < t_{\rm{dom}}$, so that the wall density remains always subdominant.  Once the volume pressure becomes comparable to the tension pressure, $p_V \simeq p_T $,  the  walls accelerate rapidly, the false vacuum disappears and the walls annihilate.
Using this condition, one can derive the annihilation time
\begin{equation}
\label{tannih} 
  t_{\rm{ann}} \simeq \dfrac{\sigma}{ V_{\rm bias}} ~,
 \end{equation}
and the corresponding annihilation temperature,
\begin{equation}
\label{Tannih}
T_{\rm{ann}} = 3  \times 10^{7}~ {\rm TeV} \left(\dfrac{10}{g_*(T_{\rm{ann}})}\right)^{1/4}
\left(\dfrac{V_{\rm bias}  }{\sigma ~ {\rm TeV}}\right)^{1/2} \simeq
3 \times 10^{7}~ {\rm TeV} \left(\dfrac{ \epsilon_b}{f_\sigma}\right)^{1/2} \left(\dfrac{v}{\rm TeV}\right)^{1/2} \, . 
\end{equation}
In the last equality we have neglected  $[g_*(T_{\rm{ann}})/10]^{1/4} = \mathcal {O}(1)$ assuming the Standard Model degrees of freedom. The requirement that  $t_{\rm{ann}} < t_{\rm{dom}}$ translates into the lower bound $V_{\rm bias} > 33 \sigma^2/ M_{Pl}^2$, or
\begin{equation}
 \label{non-domination}
\epsilon_b >  3 \times 10^{-31} \left(\dfrac{ f_\sigma\,v}{ \mathrm{TeV}}\right)^2 \, .
\end{equation}

  In order not to affect Big Bang Nucleosynthesis, we {conservatively} assume $T_{\rm{ann}} > 10$ MeV, so that the particles produced when the walls collapse have enough time to thermalize before Big Bang Nucleosynthesis. Moreover, for consistency the annihilation temperature must be smaller than the phase transition temperature, $T_{\rm{ann}} < T_c \simeq v$, which coincides with the limit in Eq.~\eqref{pVupperlimit}. Neglecting again for simplicity $[g_*(T_{\rm{ann}})/10]^{1/2} = \mathcal {O}(1)$, the allowed temperature range $10\, \mathrm{MeV} <T_{\rm{ann}} <  v$ translates into
\begin{equation}
\label{Tannih-condition}
  10^{-25} \left(\dfrac{\rm TeV}{v} \right)<\frac{\epsilon_b}{f_\sigma}<
10^{-15}\left(\dfrac{v}{\rm TeV} \right) \, .
\end{equation}

\section{Bias due to the explicit symmetry breaking}

In Section~\ref{sec:explicit}, we have introduced an explicit breaking of the flavor symmetry in the right-handed neutrino mass via the terms $\epsilon_{ij} \bar{N}_i^c N_j$, in order to generate the mixing angle $\theta_{13}$ and CP violation. These terms split the degenerate vacua with a bias which emerges at loop-level. 

Let us see the corrections to the energy of the $Z_2$ vacua due to the explicit breaking. 
For the $\chi$ flavons, the explicit breaking terms contribute to the potential of $\chi$ at one-loop,  
\begin{eqnarray}
V_{\rm loop}(\chi) = \frac{1}{64\pi^2} {\rm Tr} \left\{ \left[M_N(\chi) M_N^\dag(\chi) \right]^2 \left[\log \frac{M_N(\chi) M_N^\dag(\chi)}{\mu^2} - \frac{3}{2} \right] \right\} \,, 
\end{eqnarray} 
where $M_N(\chi)$ is the $\chi$-dependent right-handed neutrino mass matrix. In particular, $M_N(\chi)$ at $\chi = {\bm v}_{\bm 1}^+$ is fixed at $M_N$ in Eq.~\eqref{eq:mass_matrices}.
The loop correction leads to a splitting of energy of different vacua. 
Due to the special explicit breaking terms introduced in the right-handed neutrino mass matrix, $V_{\rm loop}({\bm v}_{\bm 2}^+) = V_{\rm loop}({\bm v}_{\bm 2}^-)$ and $V_{\rm loop}({\bm v}_{\bm 3}^+) = V_{\rm loop}({\bm v}_{\bm 3}^-)$.
For simplicity, in the following, we refer to the energy density difference between other vacua and ${\bm v}_{\bm 1}^+$ with $V_{{\rm bias}, \chi,j1}\equiv V_{\rm loop}({\bm v}_{\bm j}^-) - V_{\rm loop}({\bm v}_{\bm 1}^+)$, $j=1,2,3$. 
In order to ensure that ${\bm v}_{\bm 1}^+$ is the unique true vacuum, all these energy density differences have to be positive.

By taking the mass ratios of right-handed neutrinos in Eqs.~\eqref{eq:mass_ratio2}, we perturbatively obtain correlations between biases and $A_4$ explicit-breaking parameters $\epsilon_{22}$ and $\epsilon_{23}$ as
\begin{eqnarray} \label{eq:vacua_splitting}
&&(V_{{\rm bias}, \chi,21}, V_{{\rm bias}, \chi,31}, V_{{\rm bias}, \chi,11}) \nonumber\\ 
&&\simeq 10^{-3} M_N^3 v_\chi \times 
\left\{ \begin{array}{rrr}
(1.8 \epsilon_{22} - 6.3 \epsilon_{23},& -4.5 \epsilon_{22} - 6.3\epsilon_{23},& -12.6 \epsilon_{23}) \\
(2.5 \epsilon_{22} - 6.3 \epsilon_{23},& -3.9 \epsilon_{22} - 6.3\epsilon_{23},& -12.7 \epsilon_{23}) \\
(-3.8 \epsilon_{22} - 7.6 \epsilon_{23},& 3.6 \epsilon_{22} - 7.6\epsilon_{23},& -15.1 \epsilon_{23})
\end{array} \right. \, ,
\end{eqnarray}
where $\mu\simeq M_N$ has been assumed, and the top (middle) [bottom] expressions refer to the three allowed neutrino mass patterns NO1 (NO2) [IO1].

We see that the bias depends on the explicit breaking terms, in particular $\epsilon_{22}$ which is the main source of non-zero $\theta_{13}$. 
Using Eq.~\eqref{eq:theta_13}, we obtain an estimate for $\epsilon_{22}$ in terms of the heaviest right-handed neutrino mass $M_N$,  $-\epsilon_{22}\simeq \frac{M_N}{v_\chi} \times  0.16 \  (0.15) \ [-0.12] $ for the best fit value for $\theta_{13}$. 
The sign of $\epsilon_{22}$ is determined by the light neutrino mass ordering.  We recall 
correlation between $\theta_{13}$ with the right-handed neutrino masses  in Eq.~\eqref{eq:theta_13}, implying $-\epsilon_{22} \simeq \frac{M_N}{v_\chi} \times  0.16 \  (0.15) \ [-0.12] $, where we have denoted $M_N$ as the heaviest right-handed neutrino mass and we have used the best value for $\theta_{13}$, $\theta_{13} = 8.6^\circ$. Here, we have also determined the sign of $\epsilon_{22}$ based on the mass ordering. The bias is controlled also by the other explicit breaking term $\epsilon_{23}$, which is unconstrained by data. 
The requirement of having ${\bm v}_{\bm 1}^+$ as the true vacuum implies that $V_{{\rm bias}, \chi,j1}>0$, imposing restrictions on $\epsilon_{23}$. We obtain 
\begin{eqnarray}  \label{eps23}
-\epsilon_{23} > \frac{M_N}{v_\chi} \times \left\{ \begin{array}{c}  0.048 \\ 0.058 \\ 0.061 \end{array} \right. ~,
\end{eqnarray}
for NO1 (top line), NO2 (middle line), and IO1 (bottom line). Both $\epsilon_{22}$ and $\epsilon_{23}$ are of order $\sim 0.1 M_N/ v_\chi$, once we require $|\epsilon_{23}| \ll M_N/ v_\chi $. This implies that without tuned cancellations we have typically $V_{{\rm bias}, \chi} \sim {\cal O} (10^{-3} M_N^4)$.
The three biases are typically all of the same order, which as we will discuss later is relevant for observational consequences.
For instance, if we take $|\epsilon_{22}| \sim |\epsilon_{23}|$, we have $V_{{\rm bias}, \chi,21} : V_{{\rm bias}, \chi,31} : V_{{\rm bias}, \chi,11} \sim 0.35 : 0.86 : 1$ for NO1, $0.30 : 0.80 : 1$ for NO2 and $0.25 : 0.74 : 1$ for IO1. A stronger hierarchy can be obtained if one allows a fine-tuning of the parameters. In particular, if $\epsilon_{23}$ takes the smallest value allowed, Eq.~\ref{eps23}, it is possible to fine-tune $V_{{\rm bias}, \chi,21} $ to be much smaller than the other two, which are always of the same order.

It is convenient to use the dimensionless parameter $\epsilon_{b,\chi} \equiv V_{{\rm bias},\chi}/v_\chi^4$, which  is roughly estimated to be
\begin{eqnarray} \label{eq:bias_chi}
\epsilon_{b,\chi} \sim 10^{-3} \left(\frac{M_N}{v_\chi}\right)^4 \sim 10^{-3} y_N^4 \,.
\end{eqnarray}

We consider now the correction to the energy of the $Z_3$ vacua of the $\varphi$ flavons.
Since the only source of $A_4$ explicit breaking we consider comes from the right-handed neutrinos and these do not couple to $\varphi$ directly, the $Z_3$ vacua are affected much more weakly. The leading $A_4$-breaking correction to the potential of $\varphi$ is at three loops. A rough estimate is given by
\begin{eqnarray} \label{eq:varphi_bias}
V_{\rm bias, \varphi} \sim \frac{1}{(16\pi^2)^{3} }  \frac{y_{\tau}^2}{\Lambda^2} y_D^2 M_N^5 \epsilon_{ij} v_\chi = 
\frac{1}{(16\pi^2)^{3}} \frac{m_\tau^2 m_\nu M_N^7}{v_\varphi^2 v_H^4} \times \frac{\epsilon_{ij} v_\chi}{M_N}~.
\end{eqnarray}
From it, we obtain
\begin{eqnarray}
\label{Z3-corrections}
\epsilon_{b, \varphi} \equiv V_{\rm bias, \varphi}/v_\varphi^4 \sim  
10^{-27} \frac{M_N}{v_\chi}\left(\frac{M_N}{v_\varphi} \right)^6\,,
\end{eqnarray}
where $m_\nu\sim 0.1$~eV, $\epsilon_{ij} v_\chi/M_N\sim 0.1$ has been used. We see that $\epsilon_{b, \varphi}$ is much smaller than $\epsilon_{b, \chi}$, unless $v_\varphi < M_N$. 

Eqs.~\eqref{eq:bias_chi}  and \eqref{Z3-corrections} clearly imply that  the bias parameters $\epsilon_{b,\chi}$ and $\epsilon_{b, \varphi}$ can take a very large range of values, depending on the values chosen for $v_\chi$, $v_\varphi$ and $M_N$. In the next section we will show that values {of bias parameters in the approximate range 10$^{-8}$ to 10$^{-27}$, all allowed in the explicit model we consider,  could lead to observable GW signals for different values of the spontaneous-breaking scale $v$, for $v$ in the approximate range 1 TeV to $10^{11}$ TeV. Examples of models with vastly different values of these parameters will be presented in Sec.~6.}

\section{Gravitational waves: production, spectrum and limits}

The production of GW by cosmic walls was  estimated early on in~\cite{Vilenkin:1981zs, Preskill:1991kd} and later studied numerically, first in~\cite{Gleiser:1998na}  and then  in~\cite{Hiramatsu:2010yz,Kawasaki:2011vv,Hiramatsu:2013qaa} (see e.g.~\cite{Binetruy:2012ze} for a review of GW from several sources).

To estimate the  energy density in GW, we can use the quadrupole formula $P\simeq G\dddot{Q}_{ij}\dddot{Q}_{ij}$ for the gravitational power $P$ emitted by the walls  (see e.g.~\cite{Carroll:2004st,Tong_lectures}). In the scaling regime, in which $R\simeq t_U$, the quadrupole moment of the walls is $Q_{ij} \simeq E_{w} t_U^2$, where the energy in the walls is $E_{w} \simeq  \sigma R^2 \simeq \sigma t_U^2$.  Thus $\dddot{Q}_{ij} \simeq \sigma t_U$, and the power emitted in GW is  $P \simeq G \sigma^2 t_U^2$.  The energy density $\Delta \rho_{\rm gw}$ of gravitational radiation emitted in a time interval $\Delta t$ at $t_U$
is then  $\Delta \rho_{\rm gw} (t_U) \simeq P \Delta t/ t_U^3 \simeq G \sigma^2 (\Delta t/  t_U) $.  In a time interval equal to the Hubble time $\Delta t \simeq t_U$, this energy density is independent from $t_U$. This approximate estimate has been confirmed by numerical simulations~\cite{Hiramatsu:2010yz,Hiramatsu:2013qaa,Kawasaki:2011vv}. {All these simulations assume the $Z_2$ toy model presented in Sec.~3 and confirm that the simple analytic order of magnitude estimates we just presented are correct up to factors $\mathcal{O}(1)$. A better description of the GW emission in the class of models we deal with in this paper, based on more complex discrete symmetries, would require dedicated simulations.}

The contribution of the waves emitted at $t_U$ to the present-day GW energy density is redshifted by the ratio of scale factors $(a(t_U)/a_0)^4= a(t_U)^4$. Here, $a(t)$ is the scale factor of the Universe at time $t$, and we take the present scale factor to be $a_0=1$. Therefore,  the largest contribution to the GW energy density spectrum, the peak amplitude, corresponds to the latest emission time, i.e. to the time at which the walls annihilate, $t_U= t_{\rm{ann}}$.

The spectrum of GW  at time $t$ as a function of the wave-number at present $k$ (which with $a_0=1$ coincides with the comoving wave-number), or in terms of the frequency $f=k/(2\pi)$, is usually given in terms of
\begin{equation}
\Omega_{\rm gw}h^2 (k, t) = \left(\dfrac{h^2}{\rho_c(t)}\right)
 \left(\dfrac{d \rho_{\rm gw} (t)}{d \ln k}\right)~.
\end{equation}
Considering that in the scaling regime the characteristic frequency of the emitted waves is $1/t_U$ (the inverse of the typical size of the walls), the present-day frequency of waves emitted at time $t$ is $f= a(t)/t$. {Considering that $d \ln f = (H(t)- t^{-1}) dt$, for
 GW emitted in the radiation dominated period, when $H(t) = 1/2t$, we have  $d \ln f = d \ln k =-(1/2)~ d \ln t$.}  Using 
from above that
$d \rho_{\rm gw} (t) \simeq G \sigma^2 (d t/  t)$, we conclude that ${d \rho_{\rm gw} (t)}/{d \ln(k)}\simeq G \sigma^2= \sigma^2/ M_{Pl}^2$. Therefore, the peak amplitude at present is 
 \begin{equation}
\label{peak-amplitude-calc}
\Omega_{\rm gw}h^2|_{\rm peak} 
 \simeq \dfrac{h^2}{\rho_c(t_0)} \left(\dfrac{g_{s \ast}(t_0)}{g_\ast(t_{\rm{ann}})}\right)^{4/3}
\left(\dfrac{T_0}{T_{\rm{ann}}}\right)^4 \dfrac{\sigma^2}{ M_{Pl}^2}~.
\end{equation}
Here, we have used {that the GW energy density redshifts as $\Omega_{\rm gw}(t_0)= \Omega_{\rm gw}(t_{\rm{ann}}) (a(t_{\rm{ann}})/ a(t_0))^4$, the very good approximation of entropy conservation, which implies
\begin{equation}
a(t_{\rm{ann}}) = \left(\dfrac{g_{s \ast}(t_0)}{g_\ast(t_{\rm{ann}})}\right)^{1/3} \left(\dfrac{T_0}{T_{\rm{ann}}}\right) \, ,
\end{equation}
and that (except at $T<$ MeV, a range of annihilation temperatures we do not consider)  the number of entropy degrees of freedom $g_{s \ast} (T_{\rm{ann}})$ coincides with the  number of energy density degrees of freedom $g_{\ast} (T_{\rm{ann}})$. }
Using $\rho_c(t_0) = 10.53~ h^2$ keV/cm$^3$ = 8.1 $ h^2 \times 10^{-59}$ TeV$^4$  as the present critical density, $T_0 = 2.37 \times 10^{-7}$ keV as the present radiation temperature, and $g_{s \ast}(t_0)= 3.91$, we obtain
\begin{align}
\label{peak-amplitude}
\Omega_{\rm gw}h^2|_{\rm peak} & \simeq 0.7 \times 10^{-37}~\left(\dfrac{10}{g_{\ast}(T_{\rm{ann}})}\right)^{4/3}\left(\frac{\sigma}{{\rm TeV}~ T_{\rm{ann}}^2 }\right)^2  \nonumber\\
& \simeq  0.9 \times 10^{-67} \left(\dfrac{10}{g_{\ast}(T_{\rm{ann}})}\right)^{1/3}
 \left(\dfrac{\sigma}{ {\rm TeV}^3}\right)^4  \left(\dfrac{ {\rm TeV}^4}{V_{\rm bias}}\right)^2
  \simeq  1 \times 10^{-67}  \left(\dfrac{f_\sigma^4}{ \epsilon_b^2}\right) \left(\dfrac{v}{\rm TeV}\right)^4 \, .
\end{align}

The peak amplitude of the gravitational radiation spectrum  is at the frequency corresponding to the inverse of the horizon size when the walls annihilate $t_{\rm{ann}}^{-1}$, redshifted to the present,
\begin{align}
\label{peak-frequency}
f_{\rm peak}  &\simeq a(t_{\rm{ann}}) H(t_{\rm{ann}})
\simeq 1 \times 10^{-4}~ {\rm Hz} \left(\dfrac{ g_\ast(T_{\rm{ann}})}{10}\right)^{1/6} \left( \dfrac{T_{\rm ann}}{\rm TeV}\right) \nonumber\\
 & \simeq 3 \times 10^{3}~ {\rm Hz} \left(\dfrac{10}{ g_\ast(T_{\rm{ann}})}\right)^{1/12}  \left(\dfrac{V_{\rm bias}}{ \sigma \, \rm{TeV}}\right)^{1/2} \simeq   3 \times 10^{3}~ {\rm Hz} \left(\dfrac{ \epsilon_b \, v}{f_\sigma\, \rm{TeV}}\right)^{1/2} \, .
\end{align}
In the last equality, we have approximated the ratio of degrees of freedom to 1.

The relation between the peak amplitude and  the peak frequency is obtained from Eqs.~\eqref{peak-amplitude} and \eqref{peak-frequency},
\begin{align}
\label{peak-amplitude-frequency}
\Omega_{\rm gw}h^2|_{\rm peak} & \simeq 0.7 \times 10^{-17}~\left(\dfrac{10}{g_{\ast}(T_{\rm{ann}})}\right)^{2/3} \left(\dfrac{\sigma}{\rm TeV^3}\right)^2 \left(\frac{10^{-9} {\rm Hz}}{f_{\rm peak}}\right)^4~.
\end{align}
This clearly shows that walls that annihilate later, and thus produce GW with a smaller peak frequency, have a larger peak amplitude.

The simple order of magnitude estimate that we used to obtain the peak frequency is not sufficient to compute the spectrum of the GW emitted by  cosmic walls. This spectrum has been computed analytically and numerically  for the simple toy model in Eq.~\eqref{toy-model}~\cite{Hiramatsu:2013qaa}. It has a $k^3$ dependence for $k < k_{\rm peak}$  and a  $1/k$ dependence for $k > k_{\rm peak}$.  Waves with frequency below the peak correspond to  super-horizon wavelengths at $t_{\rm{ann}}$. Causality requires the spectrum to go to zero as $k^3$. Indeed, this is a characteristic of a white noise spectrum as it corresponds to the absence of causal correlations~\cite{Caprini:2009fx}. The spectrum at frequencies above the peak depends instead on the particular model assumed to produce the waves. The spectrum $1/k$ was found analytically for a source that is not correlated at different times, i.e. that consists of a series of short events~\cite{Caprini:2009fx} and also numerically for the toy model we consider. Contrary to previous studies finding a flat spectrum at large $k$ \cite{Hiramatsu:2010yz,Kawasaki:2011vv}, the $1/k$ spectrum was eventually confirmed also with numerical simulations~\cite{Hiramatsu:2013qaa}.

\begin{figure}{
\begin{center}
\begin{subfigure}{.5\textwidth}
  \centering
  \includegraphics[width=1\linewidth]{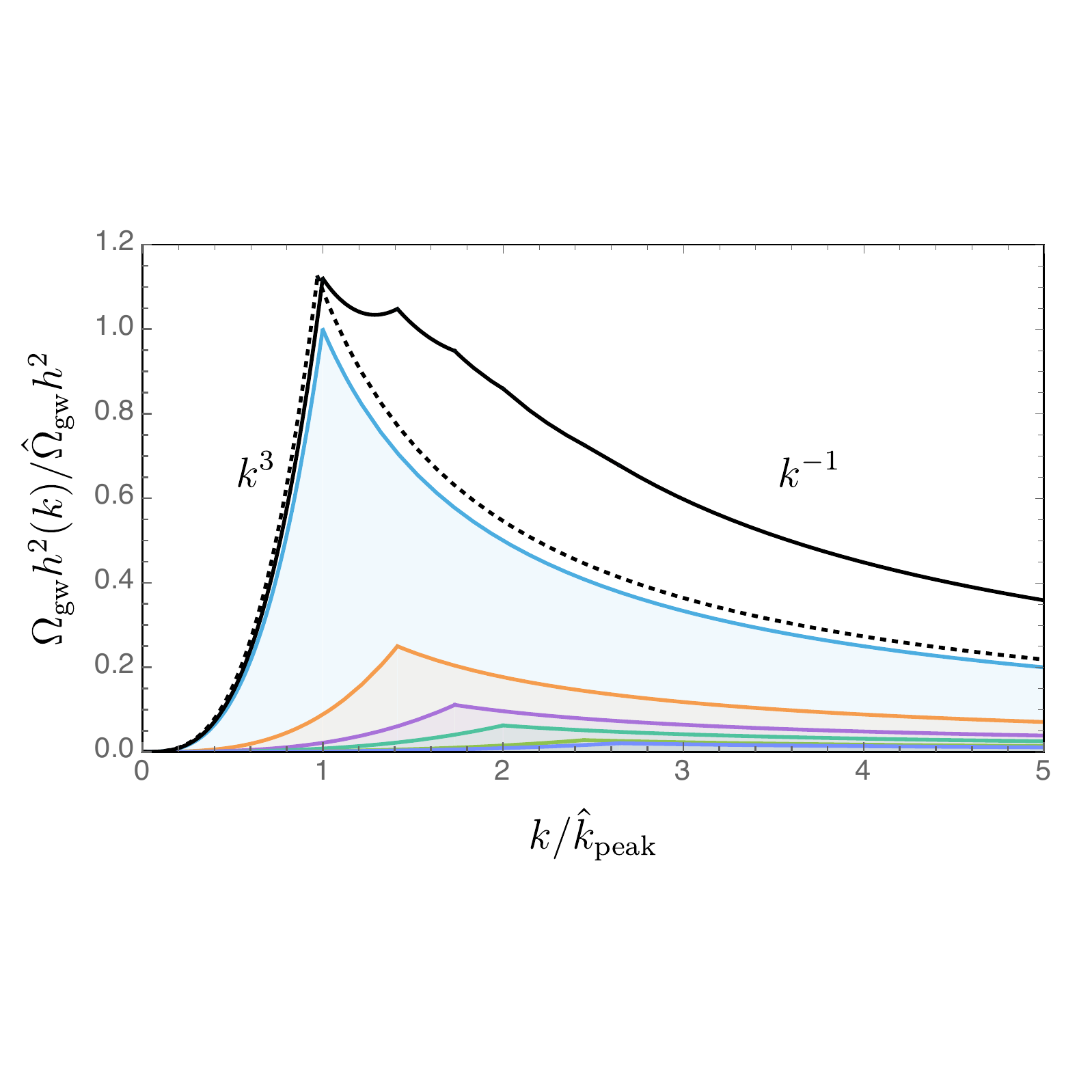}
\end{subfigure}%
\begin{subfigure}{.5\textwidth}
  \centering
  \includegraphics[width=1\linewidth]{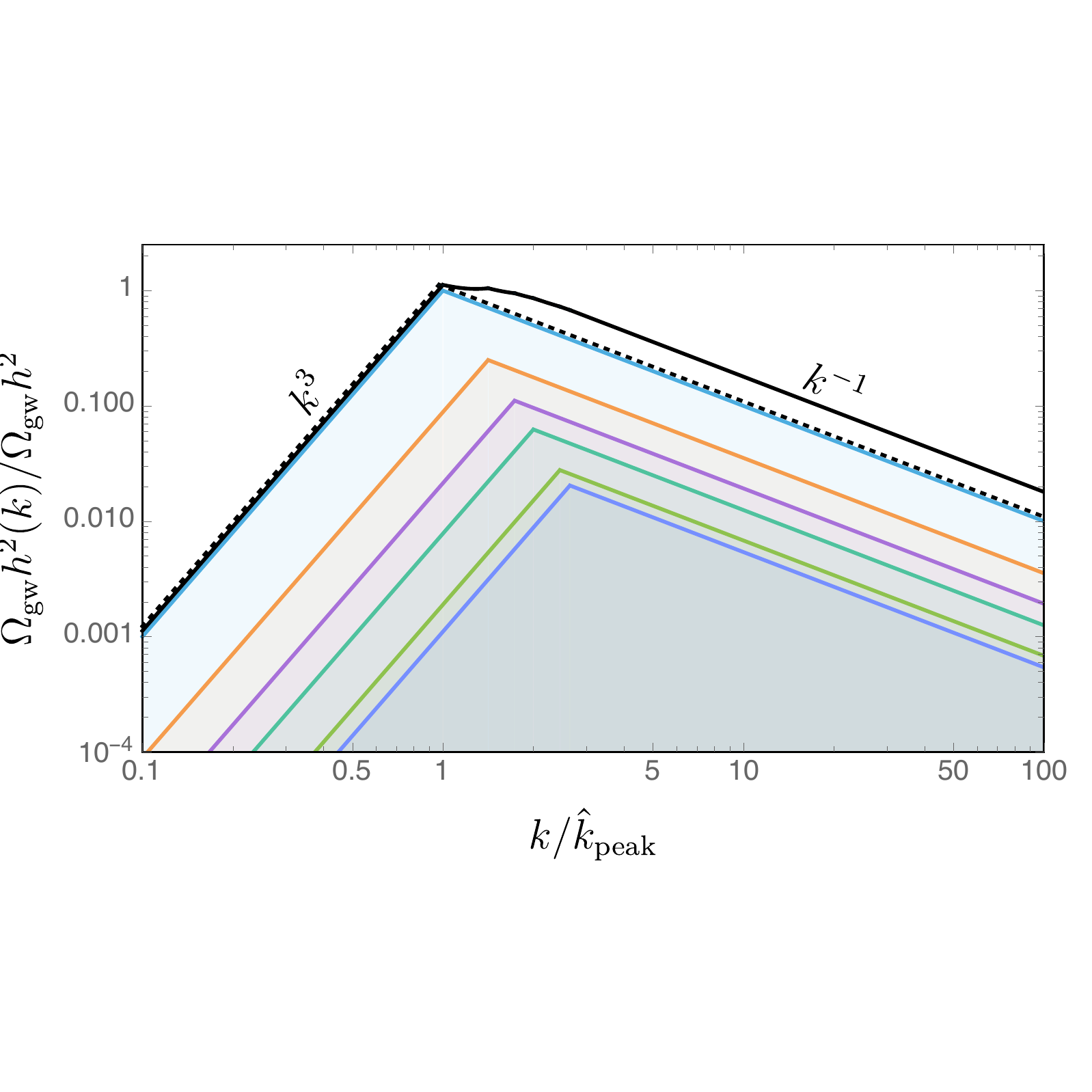}
\end{subfigure}
\end{center}
\vspace{-5em}
\caption{Gravitational waves spectrum due to overlapped spectra, presented in linear (left) and logarithmic (right) scales, each corresponding to populations of walls. We show six different values of the bias potential between two vacua, $V_\mathrm{bias}=\{1,2,3,4,6,7\}~ \hat{V}_\mathrm{bias}$.  
These arise from the energy differences of four vacua, with energy $V=\{0,1,3,7\}~\hat{V}_\mathrm{bias}$.
The wave-numbers and the spectrum are respectively normalized to $\hat{k}_\mathrm{peak}\sim{V}_\mathrm{bias}^{1/2}= \hat{V}_\mathrm{bias}^{1/2}$, and its corresponding amplitude $\hat{\Omega}_{\rm gw}h^2$. The spectrum generated by walls with a single bias value and peak amplitude equal to the total spectrum one is also shown (dashed black), to highlight the difference of this case with our scenario in which multiple values of the bias are expected.
\label{spectrafig}}}
\end{figure}

As shown in Eq.~\eqref{eq:vacua_splitting}, in the case of flavons generically there are
several values of $V_{\rm bias}$. In our particular example,  $V_{\rm bias}$ for the $\chi$ flavons take several values, that are within a factor of a few of each other, and the same happens for the $\varphi$ flavons. These two sets of values are typically separated by many orders of magnitude. This results in separate sets of walls,  each of which would produce GW with  multiple close peaks, as shown in  
Fig.~\ref{spectrafig}. We show the toy case of four different vacua, giving rise to six peaks in the GW spectrum.  With a 
general breaking, there could be up to fifteen different values of $V_{\rm bias}$ among the six different vacua of the $\chi$ flavons 
and possibly many more among the $\varphi$ flavons, all {of the same order of magnitude within factors of a few}
of each other. In general, for $n$ degenerate vacua there could be up to $n!/[2 (n-2)!]$ different values of the bias. Therefore, observing GW with multiple peaks very close in frequency is a generic feature of flavon generated walls. Our discussion is mostly qualitative {and necessarily disregards cross terms in the energy momentum tensor which would soften the transition between peaks}. In order to predict the spectrum in a more realistic manner, a numerical study of the wall network in the models we consider, with discrete symmetries more complicated than $Z_2$, is needed. We leave such a study for future work. 

\begin{figure}[htb]
\begin{center}
\includegraphics[width=0.9
\textwidth]{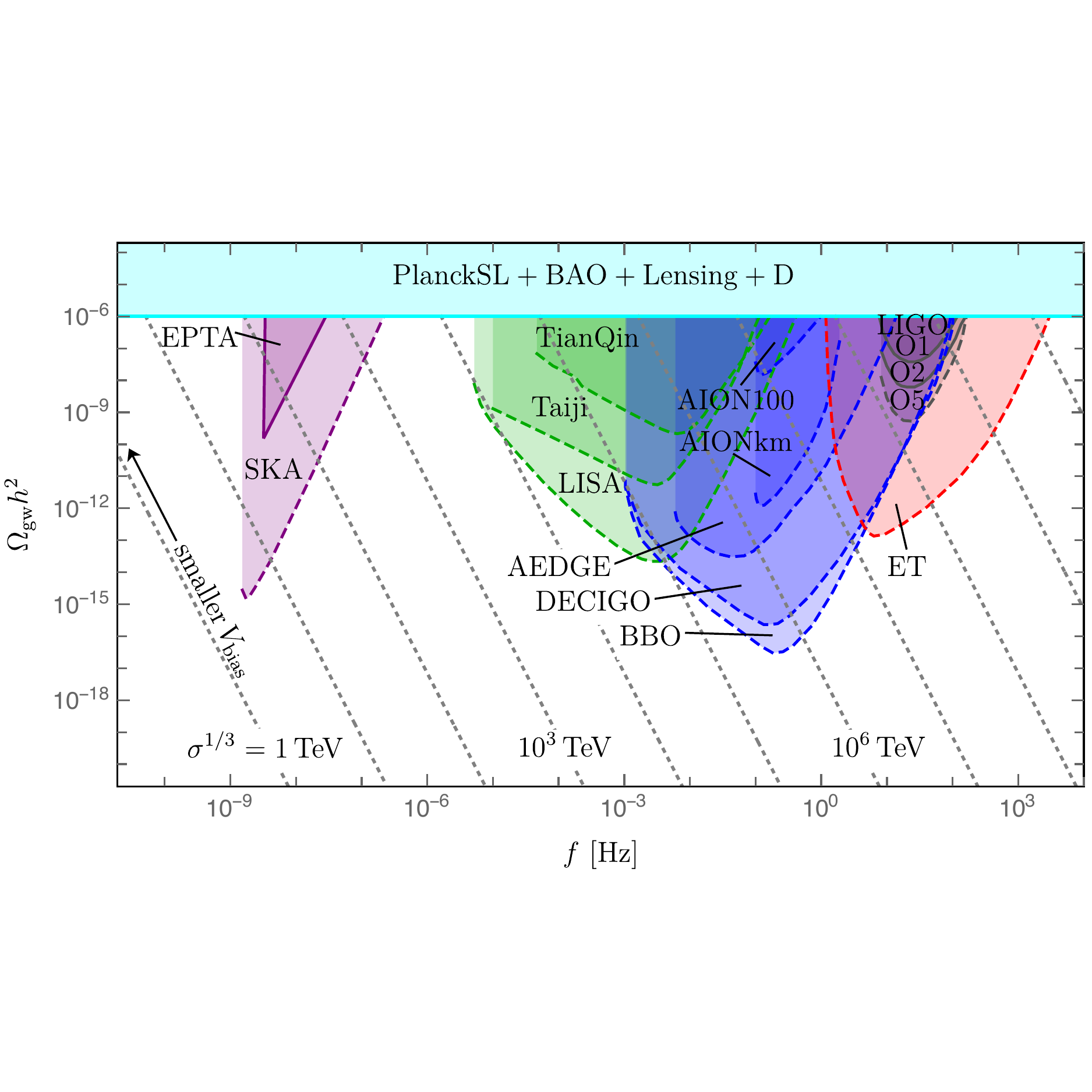}\vspace{-5em}
\caption{ 
\label{fig:GW-data}  Lines of $f_\mathrm{peak}$ for fixed values of $\sigma$ (dotted black), from Eq.~\eqref{peak-frequency} compared to 95\% C.L. upper limits (solid lines) and planned sensitivity reach (dashed lines)  of several GW experiments.
We show: radio telescopes in purple, the European Pulsar Timing Array (EPTA) \cite{Lentati:2015qwp} and the Square Kilometre Array (SKA) \cite{Janssen:2014dka} limit and reach, respectively,  the projected sensitivities of the space-based experiments TianQin \cite{Luo:2015ght}, Taiji \cite{Guo:2018npi}, and the Laser Interferometer Space Antenna (LISA) \cite{Audley:2017drz} in green, the reach of the Atom Interferometer Observatory and Network (AION) \cite{Badurina:2019hst}, the Atomic Experiment for Dark Matter and Gravity Exploration in Space (AEDGE)  \cite{Bertoldi:2019tck}, the Deci-hertz Interferometer Gravitational wave Observatory (DECIGO) \cite{Seto:2001qf}, and the Big Bang Observer (BBO) \cite{Corbin:2005ny} in blue. Finally, we show in red the reach of the ground based experiments Einstein Telescope (ET) (projection) \cite{Sathyaprakash:2012jk} and in grey limits and future reach of the Laser Interferometer Gravitational-Wave Observatory (LIGO) \cite{LIGOScientific:2019vic}.The cyan band is rejected by the 95\% CL upper limit on the effective number of degrees of freedom during CMB emission, from Planck and other data~\cite{Pagano:2015hma}. }
\end{center}
\end{figure}

\begin{figure}[htb]
\begin{center}
\includegraphics[width=.90
\textwidth]{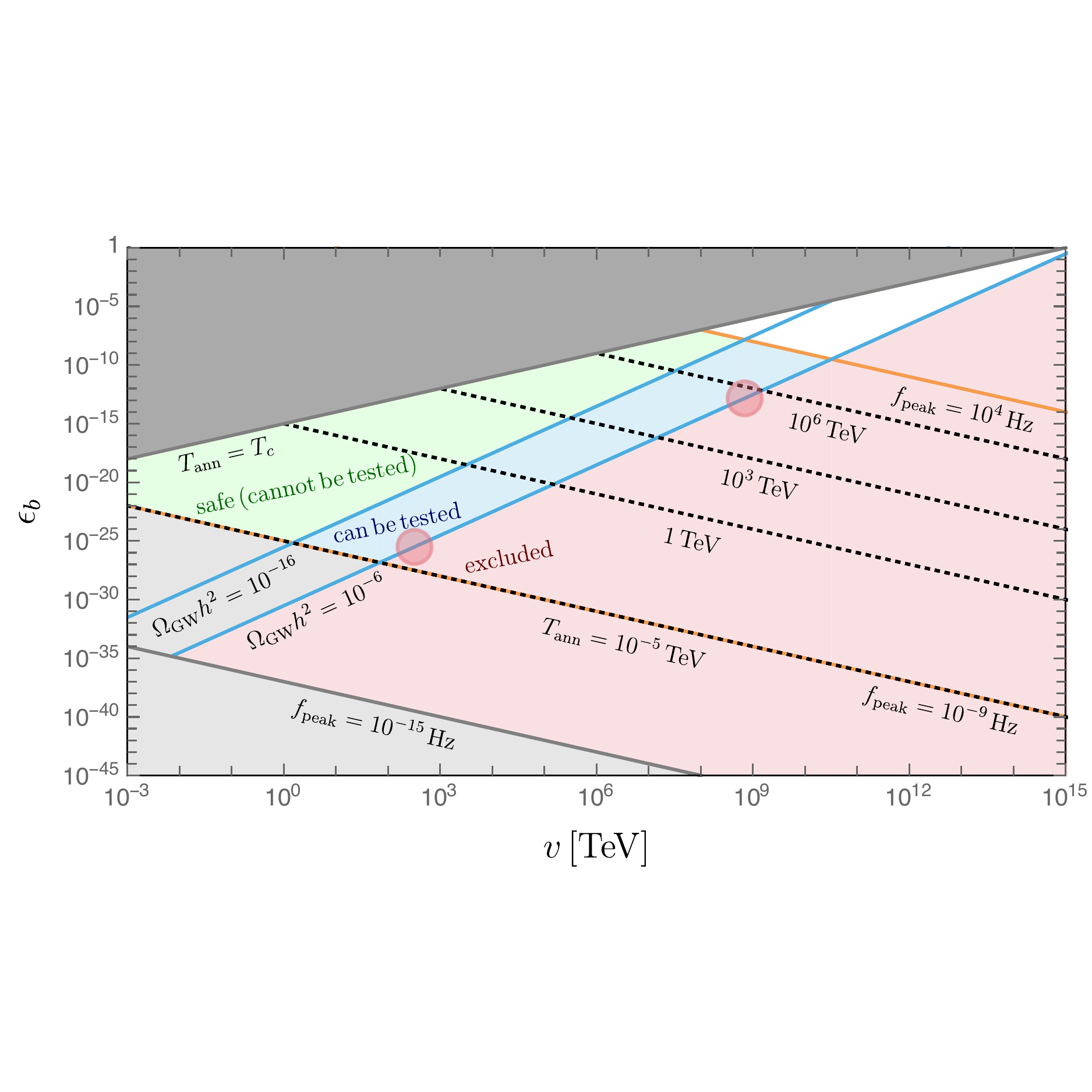}\vspace{-5em}
\caption{  \label{fig:Main}
Regions of interest of the dimensionless bias potential parameter $\epsilon_b$ in Eq.~\eqref{bias} as function  of the spontaneous symmetry breaking scale $v$ derived from the peak density and peak frequency in Eqs.~\eqref{peak-amplitude} and~\eqref{peak-frequency} and assuming $f_\sigma=1$. We show where walls are forbidden by present upper limits on gravitational radiation and where  walls would dominate the energy density of the Universe (red) and where GW from wall annihilation are allowed and either will be observable (blue and part of white) or not  (green) in future GW experiments (see Fig.~\ref{fig:GW-data}). The red circles show approximately regions excluded experimentally by EPTA and LIGO. No walls can exist in the upper gray
triangular region and we have not extended our analysis to the light gray regions in the lower left corner. See the text for details. 
}
\end{center}
\end{figure}

We use the peak frequency and amplitude to explore the observability of the GW signal produced by cosmic walls, thus we are not going to rely on the specific spectral shape. The existing limits and future reach of GW detectors are shown in Fig.~\ref{fig:GW-data}.   Our main results are given in Fig.~\ref{fig:Main}, where we show the allowed ranges of the bias potential as function of the spontaneous symmetry breaking scale $v$. The bias is parametrized through the dimensionless parameter $\epsilon_b$, defined in Eq.~\eqref{bias}. To produce this figure we used the peak density and peak frequency given in Eqs.~\eqref{peak-amplitude} and~\eqref{peak-frequency}, assuming $f_\sigma=1$, and we constrained them using Fig.~\ref{fig:GW-data}. Notice that other
values of $f_\sigma$ would change the regions shown in Fig.~\ref{fig:Main}. In our paradigm $A_4$
model $f_\sigma$ is generally of order one.

No walls can form within the upper grey triangular region of Fig.~\ref{fig:Main} (where walls would annihilate before they exist). Only below this grey region the annihilation temperature is smaller than the critical temperature at which walls form, $T_{\rm annih} < T_c$. Lines of constant $T_{\rm annih}$ (dotted black)  are shown for temperatures between 10 MeV and $10^8$ TeV, a range that through Eq.~\eqref{peak-frequency} corresponds to observable peak frequencies, between $10^{-9}$ and $10^4$ Hz, see Fig.~\ref{fig:GW-data}. As shown in Fig.~\ref{fig:GW-data}, in this range of frequencies, GW
will be observable if $\Omega_{\rm gw}h^2$ is roughly between $10^{-6}$ and $10^{-16}$, i.e. within the blue region in Fig.~\ref{fig:Main}. 

 If they are observed, GW emitted by the annihilation of flavon cosmic walls {could possibly} be identified by the characteristic multipeaked spectrum of the type shown in Fig.~\ref{spectrafig}. This spectrum is composed of several overlapped single spectra, with frequency dependence $f^3$ below and $f^{-1}$ above the peak $f_{\rm peak}$, with peaks close in frequency, {all of the same order of magnitude, within a factor of $\mathcal{O}(1)$ of each other}, each with peak amplitude 
$\Omega_{\rm gw}h^2|_{\rm peak} \sim f_{\rm peak}^{-4}$, as written in Eq.~\eqref{peak-amplitude-frequency}. {This prediction disregards the effect of crossed terms in the source energy momentum tensor, which would soften the transition between peaks, and could only be taken into account with a numerical simulation of the evolution of walls in the models we consider.} The blue region in Fig.~\ref{spectrafig}
shows where the peak frequency and amplitude are within the observable range, but GW could also be detected by the tail of the spectrum below the maximum frequency. Thus, part of the white region above the blue one, where $f_{\rm peak}> 10^4$ Hz (i.e., $T_{\rm annih} > 10^8$ TeV), should be observable by ET, and maybe also LIGO, DECIGO and BBO, through the lower frequency $f^3$ dependent part of the spectrum (we do not attempt here to quantify this effect). 

The upper limit $\Omega_{\rm gw} h^2 < 1 \times 10^{-6}$ stemming from the 95\% CL upper limit on the effective number of degrees of freedom during CMB emission from Planck and other data~\cite{Pagano:2015hma} exclude the region shown in red in the figure.  This upper limit applies only to GW frequencies above $10^{-15}$ Hz, which corresponds to the inverse horizon at recombination. We do not consider any limit at lower frequencies, which would apply in the light grey lower right corner of the figure. Future observation could tighten the CMB upper limt on $\Omega_{\rm gw} h^2$ by up to one order of magnitude~\cite{Pagano:2015hma}.  The red circles indicate approximately the region of parameter space where there are already experimental exclusion limits imposed by EPTA and LIGO (see the solid line contours in Fig.~\ref{fig:GW-data}).

By coincidence, the red region in Fig.~\ref{fig:Main} is also rejected by the requirement that walls do not get to dominate the energy density of the Universe, $t_{\rm{ann}} < t_{\rm{dom}}$, in Eq.~\eqref{non-domination}.

The green region in  Fig.~\ref{fig:Main} corresponds to
GW amplitudes $\Omega_{\rm gw} h^2 < 1 \times 10^{-16}$ which could not be reached by any of the experiments in Fig.~\ref{fig:GW-data} and correspond to walls that annihilate sufficiently before BBN, at $T_{\rm annih} > 10$ MeV,  as to allow for the thermalization of the particles produced by the annihilation before BBN. Extending this ``safe" green region as well as the ``observable" blue region to $T_{\rm annih} < 10$ MeV, to the light grey region, would imply studying the effects of the annihilation of walls during and after BBN, which is outside the scope of this paper.

\section{Gravitational waves from discrete flavor symmetries}

Through a judicious choice of parameters in the models we consider, the spontaneous symmetry breaking of $A_4$ and bias due to its explicit breaking would lead to observable GW, or otherwise to cosmologically viable models in which the GW produced are not intense enough to be observable in the foreseeable future. We show this specifically in our $A_4$ model for both the $\chi$ and $\varphi$  flavons.

We see from Eq.~\eqref{eq:bias_chi} that the size of $\epsilon_{b,\chi}$ is directly determined by the Yukawa coupling $y_N$, which is of order $\mathcal{O}(M_N/v_\chi)$. Once $y_N$ (and consequently $\epsilon_{b,\chi}$) is fixed, $v_\chi$ can vary in a very wide range depending on the right-handed neutrino mass range. This feature can provide a potential connection with other phenomenological studies in particle physics and cosmology.  A particularly interesting link arises with leptogenesis. This process requires right-handed neutrino with masses which can go from 1 to $10^{15}$~GeV, depending on the specific implementation used. In Table~\ref{tab:benchmarks}, we list some benchmark points of particular interests, which are consistent with current cosmological constraints and can be tested in future GW observatories, see Fig.~\ref{fig:Main}. 
B1 corresponds to the spontaneous breaking of the flavor symmetry at the GUT scale $v_\chi \sim 10^{13}$ TeV, and for $y_D\sim 0.1$ right-handed neutrinos with masses $M_N \sim 10^{12}$ TeV, which is the typical see-saw scale and one at which leptogenesis proceeds usually in the one-flavor approximation.
For smaller masses (e.g. B2), flavor effects in leptogenesis start becoming important \cite{Davidson:2008bu} as (some of) the charged leptons get in equilibrium in the early Universe. For lower scales, such as in B3 and B4, flavor effects are fully developed. Lowering the scale further, as in B5, would allow to search directly for the right-handed neutrinos at colliders.
Finally, $M_N \sim$ GeV in B6 refers to the mass scale accessible at peak and heavy neutral lepton decay searches.
We do not consider lower right-handed neutrino mass due to the strong constraints from Big Bang Nucleosynthesis~\cite{Ruchayskiy:2012si,Gelmini:2020ekg}.

\begin{table}[ht]
\centering
\begin{tabular}{c c c c c c}
\hline\hline\\[-12pt]
Benchmark & $y_N$ & $M_N/{\rm TeV}$ & $v_\chi /{\rm TeV}$ & $\epsilon_{b,\chi}$ & Testability \\ [2pt] 
\hline\\[-12pt]
B1& $10^{-1}$ & $10^{12}$ & $10^{13}$ & $10^{-7}$ & Excluded \\
B2& $10^{-1}$ & $10^{9}$ & $10^{10}$ & $10^{-7}$ & Not observable \\
B3& $10^{-2}$ & $10^{6}$ & $10^{8}$ & $10^{-11}$ & Observable \\
B4& $10^{-3}$ & $10^{3}$ & $10^{6}$ & $10^{-15}$ & Observable \\
B5& $10^{-4}$ & $10^0$ & $10^{4}$ & $10^{-19}$ & Observable \\
B6& $10^{-5}$ & $10^{-3}$ & $10^2$ & $10^{-23}$ & Observable \\
B7& $10^{-4}$ & $10^{-3}$ & $10^1$ & $10^{-19}$ & Not observable \\
\hline
\end{tabular}
\caption{Benchmark points for $y_N$ and $\chi$ bias  (order of magnitude). The correlation $y_N \simeq M_N/v_\chi$ is explicitly given. }
\label{tab:benchmarks}
\end{table}

Using Eq.~\eqref{Z3-corrections} for the $\varphi$ flavon bias generated at three-loop order, we list in Table~\ref{tab:benchmarks2} some benchmarks values of $\epsilon_{b,\varphi}$ in the case of $v_\varphi \sim v_\chi$. All these benchmarks give very tiny biases, which lead to  domain walls  that collapse too late (cf. Eq.~\eqref{Tannih}) to be allowed  by cosmological limits. 

\begin{table}[ht]
\centering
\begin{tabular}{c c c c c c }
\hline\hline\\[-12pt]
Benchmark & $M_N/{\rm TeV}$ & $v_\chi /{\rm TeV}$ & $\epsilon_{b,\varphi}$ ($v_\varphi \simeq v_\chi$) & $\epsilon_{b,\varphi}$ ($v_\varphi \simeq 10^{-4} v_\chi$) & Testability \\ [2pt] 
\hline\\[-12pt]
B1$'$& $10^{12}$ & $10^{13}$ & $10^{-34}$ & $10^{-7}$ &  Not observable\\
B2$'$& $10^{9}$ & $10^{10}$ & $10^{-34}$ & $10^{-10}$ & Not observable \\
B3$'$& $10^{6}$ & $10^{8}$ & $10^{-41}$ & $10^{-17}$ & Observable \\
B4$'$& $10^{3}$ & $10^{6}$ & $10^{-48}$ & $10^{-24}$ & Observable \\
B5$'$& $10^0$ & $10^{4}$ & $10^{-55}$ & $10^{-31}$ & Not considered \\
B6$'$& $10^{-3}$ & $10^2$ & $10^{-62}$ & $10^{-38}$ & Excluded \\
B7$'$& $10^{-3}$ & $10^1$ & $10^{-55}$ & $10^{-31}$ & Not considered \\
\hline
\end{tabular}
\caption{Benchmarks points for $\varphi$ bias in the charged lepton sector for different values of $M_N$ and $v_\chi$. Two sets for the VEV $v_\varphi$, $v_\varphi \simeq v_\chi$ and $v_\varphi \simeq 10^{-4} v_\chi$ are considered. Results are provided as order of magnitude estimates. All benchmarks for the set $v_\varphi \simeq v_\chi$ have been excluded. Testability for the set $v_\varphi \simeq 10^{-4} v_\chi$ is shown in the table. The label ``Not considered" means that the model is in a part of the parameter space not covered by our calculations (the gray regions of Fig~4). 
}
\label{tab:benchmarks2}
\end{table}

There are several possible ways to avoid this problem. The first is to assume
    a period of inflation such that the $\varphi$ domain walls are inflated away by inflation and the $A_4$ symmetry later is not restored by reheating processes at the end of inflation~\cite{Riva:2010jm, Antusch:2013toa}.
    Note that if we still hope to observe GW signatures from collapsing domain walls in this case, $\varphi$ and $\chi$ cannot gain VEV at the same time. A time ordering of the main events is required, in which the $\varphi$ walls formed first, inflation happens later or at the same time, and then the $\chi$ walls in the end.

    A way to make the model compatible with cosmology and to potentially detect gravitational waves from the $\varphi$ flavon sector is to assume a hierarchical VEV ordering $v_\varphi \ll v_\chi$. In the last column of 
    Table~\ref{tab:benchmarks2}, we modified the benchmarks by fixing $v_\varphi=10^{-4} v_\chi$.
    In this case most benchmarks predict $\varphi$ domain walls consistent with the standard cosmology except B6$'$. We notice that the B6$'$ and B7$'$ choices have too small $\varphi$ VEV, $v_\varphi \sim$ 10 GeV and 1 GeV, and thus are not consistent with constraints from charged lepton flavor violation (CLFV) \cite{Pascoli:2016wlt} and LHC searches \cite{Heinrich:2018nip}. B5$'$ assumes $v_\varphi$ at the TeV scale. This benchmark would be testable in future CLFV measurement and collider searches. 
    
    The last possibility we envision to avoid cosmological problems consists in including
    additional sources of explicit breaking. In analogy to the terms we added to right-handed neutrino mass terms, one may include explicit breaking mass terms for charged vector-like leptons $E_i$. For example, one could add  small $A_4$-breaking terms $\epsilon'_{ij} M_E \bar{E}_{iL} E_{jR}$ (with $|\epsilon'_{ij}| \ll 1$) to the mass term $M_E \bar{E}_{iL} E_{iR}$. 
    
  Other possible  sources of explicit breaking to solve the domain problem have been discussed in the literature. For example,    by extending $A_4$ to the quark sector, the $A_4$ symmetry could gain anomalous loop corrections due to the QCD anomaly~\cite{Preskill:1991kd}. Recently,~\cite{Chigusa:2018hhl} discussed in detail that, by arranging quark flavors as singlets of $A_4$, the degeneracy of the $Z_3$ vacua can be removed, but the degeneracy of true vacuum can only be partially removed, thus the walls problem is not solved. However, including the QCD anomaly does not split the degeneracy of $Z_2$ vacua, and thus, it cannot fully solve the domain wall problem.

\section{Comment on walls evolution with viscosity}
\label{viscosity}

As explored originally by Kibble~\cite{Kibble:1976sj}, a viscosity pressure  can appear due to a frictional force generated by the reflection on the walls of particles constituting the thermal bath (see also~\cite{Vilenkin:2000jqa}). A large friction force of the thermal bath on the walls exists when a relativistic mass eigenstate on one side corresponds to a superposition of mass eigenstates on the other side, and a significant component of it is a non-relativistic state~\cite{Everett:1974bs}.

The reflection and transmissivity depend, among other factors, on the de Broglie wavelength of the scattering particles compared to the thickness of the wall. As in a thermal bath relativistic particles have a typical momentum $k\simeq T$, to study their reflection or transmission the walls can be safely considered thin, i.e. $k\lesssim 1/\Delta\simeq v$ (see the wall thickness $\Delta$ in Eq.~\eqref{wall-thickness} for the toy model in Eq.~\eqref{toy-model}). In the thin-wall approximation, a total or partial reflection of relativistic particles can happen when their mass eigenstates on the two sides of the wall are different, i.e. when the matrix mixing the interaction and mass eigenstates is different in different vacua.  Assume for simplicity just two mass eigenstates with mass $m$ and $M$, where $m \ll T\ll M$, and they are rotated by an angle $\alpha> m/M$ from one side to the other. 

When a relativistic particle with momentum $k$ moving along the negative $x$ axis collides with
a wall at $x=0$, it can be treated as a plane wave of amplitude $A_0$ which is partially reflected (with amplitudes $A_R$ and $B_R$ for the light and heavy states) in the initial side
\begin{align}
    A_0\begin{pmatrix} 1 \\ 0 \end{pmatrix}e^{ikx}+ A_R \begin{pmatrix} 1 \\ 0 \end{pmatrix} e^{-ikx}+B_R\begin{pmatrix} 0 \\ 1 \end{pmatrix} e^{px}~, 
\end{align}
and partially transmitted (with amplitudes $A_T$ and $B_T$) to the vacuum on the other side of the wall,
\begin{align}
    A_T\begin{pmatrix} \cos\alpha \\ \sin\alpha \end{pmatrix}e^{ikx}+ B_T \begin{pmatrix} -\sin\alpha \\ \cos\alpha \end{pmatrix}e^{-px}.
\end{align}
Here, $p=\sqrt{M^2-k^2}$  (energy is conserved but momentum is not as the wall breaks translation symmetry). The different amplitudes are determined
by imposing the continuity of the solution and its first derivative at the wall, $x=0$. When $M\gg T$ the solutions are
$|A_R|\simeq |A_0|$, $|B_R|\simeq |A_T|\simeq |B_T|\simeq 0$, i.e. we have total reflection. The general expression for the reflected wave amplitude is
\begin{equation}
\frac{A_R}{A_0}=	\frac{\sin^2{\alpha}(k^2+m^2-M^2-p^2)}{4  \cos^2{\alpha}\sqrt{k^2+m^2}\sqrt{p^2+M^2}+\sin^2{\alpha}(k^2+m^2+p^2+M^2+2\sqrt{k^2+m^2}\sqrt{p^2+M^2})}
\end{equation}
which goes to zero if  $\sin{\alpha}$  goes to zero, and for  $\sin{\alpha} >  k/M \simeq T/m > m/M$ in the regime in which $m \ll k \ll M$ where reflection may be important
\begin{equation}
\left|\frac{A_R}{A_0}\right|^2 \simeq \left(1 + \frac{\sqrt{2}\, k}{M}\right)^{-2}  
\end{equation}
which goes to 1 for $M\gg k$.

The viscosity pressure on a wall moving with speed $v_{w}$ with respect to the thermal bath  can be written as~\cite{Kibble:1976sj}
\begin{equation}
\label{viscosity-pressure}
p_{\rm visc} \simeq \alpha \rho_{r}  v_{w}  \simeq \alpha   v_{w} T^4,
\end{equation}
where the real coefficient $\alpha$ is $0\leq \alpha<1$. The latter is  close to 1 if at least one of the relativistic species in the thermal bath is strongly reflected when colliding with the walls, so that the momentum transferred per collision is approximately  $T$.

The presence of viscosity alters considerably the evolution of the walls with respect to the scenario we described earlier (see e.g.~\cite{Preskill:1991kd}). In the absence of viscosity walls straighten out to the horizon scale shortly after they form, even if the initial correlation length is smaller than the horizon.  With viscosity, only the wall features at scales much smaller than the
horizon straighten out, so the characteristic linear scale is $R_{\rm smooth} \ll H^{-1}$. Moreover, in the presence of a  non-negligible bias, which tends to accelerate the walls towards the false vacuum regions, the main pressure opposing the acceleration  would now be due to viscosity. We therefore estimate the annihilation to happen when the
pressure due to bias overcomes the viscosity resistance and $ v_{w}$ becomes close to 1, so that
\begin{equation}
\label{Tannih-visc}
p_V \simeq V_{\rm bias}= \epsilon_b v^4 \simeq p_{\rm visc} \simeq \alpha T_{\rm{ann}}^4  \, .
\end{equation}

This type of viscosity dominated evolution could happen due to the reflection of flavons shortly after the phase transition. Flavons acquire a mass of order $v$, but flavon masses may differ by a few orders of magnitude, so the lightest flavon could have mass $\geq 10^{-3} v$. 

Let us consider as an example the three flavons $\chi_1$, $\chi_2$ and $\chi_3$. Given the flavon potential in Eq.~\eqref{eq:self-coupling1}, they gain masses $M_{\chi_1} = \sqrt{2 g_1}\, v_\chi$, $M_{\chi_2} = M_{\chi_3} = \sqrt{g_2/2}\, v_\chi$ in the vacuum ${\bm v}_{\bm 1}^\pm$, $M_{\chi_2} = \sqrt{2 g_1}\, v_\chi$, $M_{\chi_1} = M_{\chi_3} = \sqrt{g_2/2}\, v_\chi$ in ${\bm v}_{\bm 2}^\pm$, and $M_{\chi_3} = \sqrt{2 g_1}\, v_\chi$, $M_{\chi_1} = M_{\chi_2} = \sqrt{g_2/2}\, v_\chi$ in ${\bm v}_{\bm 3}^\pm$ (see Appendix~\ref{sec:B}). Notice that while the flavon masses are the same in all vacua (if the symmetry is exact), the composition of the flavon mass eigenstates in terms of the interaction eigenstates is different in different vacua. For example, at walls separating ${\bm v}_{\bm 1}^+$ from ${\bm v}_{\bm 2}^\pm$, the mixing angle  between $\chi_1$ and $\chi_2$ is $\alpha = 90^\circ$. We have the same mixing angle for transmitting between $\chi_1$ and $\chi_3$ at walls separating ${\bm v}_{\bm 1}^+$ from ${\bm v}_{\bm 3}^\pm$.

In the extreme case in which some couplings in the flavon potential are hierarchical, e.g., $g_1 \sim 1$ and $g_2 \sim 10^{-6}$, a hierachical mass spectrum for flavons may be generated, in which the lightest flavon has a mass 10$^{-3}$ of the heaviest. So, while the lightest flavon is relativistic and the heaviest is non-relativistic, there could be a strong reflection of the lightest flavons by the walls. Since this  can only happen for a few orders of magnitude  in temperature below the critical temperature, a viscosity dominated evolution can only happen if the bias is large enough, such as $T_{\rm{ann}} > 10^{-3} T_c \simeq 10^{-3} v$.  Using Eq.~\eqref{Tannih-visc}, this would require a very large bias, with 
\begin{equation}
\label{bias-visc}
\epsilon_b  > 10^{-12} \, .
\end{equation}
While this is achievable in the models we consider, the GW produced during the phase transition itself could not be neglected, since the walls annihilate shortly after the phase transition.  This is a very complicated and interesting scenario which may require detailed simulations and we thus do not pursue any further in this paper.

\section{Conclusions}

We have shown that lepton mass and mixing models based on
spontaneously broken discrete symmetries can lead to characteristic gravitational wave signatures, produced by the annihilation of the cosmic walls which appear in these models. 
So far, most of them assumed that the spontaneous breaking happens before inflation,  so that walls are inflated away.
We showed that 
the cosmic walls can be cosmologically safe for any symmetry breaking energy scale below $M_{ Pl}$ without
the need to resort to inflation, with the introduction of an adequate explicit symmetry breaking. To illustrate the main features of this scenario, we  used a specific realization based on the tetrahedral group $A_4$.

These flavor discrete symmetries necessitate several scalar fields, the flavons, whose vacuum expectation values produce the desired spontaneous breaking.
This leads to many  equivalent minima of the flavon potential, all degenerate if the discrete symmetry is exact. However, neutrino oscillation data require modifications of the leading order predictions, which can be accomplished with an explicit breaking. For example, an explicit symmetry breaking of the $A_4$ symmetry  can account for the non-zero value of the $\theta_{13}$ neutrino mixing angle. This explicit breaking can also lift the degeneracy of the multiple minima of the scalar potential. The energy difference between different vacua, called ``bias'',  drives the evolution of the walls towards their annihilation. The bulk of the gravitational waves is generated when walls annihilate.  We have parametrized the bias  as $V_{\rm bias} = \epsilon_b \, v^4$, where $v$ is the spontaneous symmetry breaking scale. Regions of interest in the $v$, $\epsilon_b$ plane are shown in Fig.~\ref{fig:Main}. Generically, with an explicit breaking in the right-handed neutrino sector, neutrino oscillation data do not severely constrain
the order of magnitude of the bias, which appears at loop order.

An explicit breaking of a non-Abelian discrete symmetry generates bias values between the multiple vacua. The latter are naturally all close to each other, within one order of magnitude. In our $A_4$ example, the split between  the right-handed neutrino mass scale and $v$ suppresses the differences between different bias values. This produces a distinctive spectrum  given by the sum of several overlapped spectra, one for each bias, that have peak frequencies $f_{\rm peak}\sim V_\mathrm{bias}^{1/2}$ {of the same order of magnitude} and maximum amplitudes  proportional to  $f_{\rm peak}^{-4}$ (as shown in Fig.~\ref{spectrafig}). {Crossed terms in the source energy-momentum tensor, that we necessarily disregard, would soften the transition between peaks. This effect could only be taken into account in numerical studies of the evolution of the wall system in the models we consider. We leave such a study for future work.}

Figure~\ref{fig:Main} shows that cosmic walls from  non-Abelian discrete flavor symmetries can be cosmologically safe (i.e. within the blue, green, or white regions in the figure) for any spontaneous symmetry breaking scale $v$ between 1 and  $10^{18}$ GeV, if the bias parameter is chosen adequately. The choice of bias in turn provides constraints on the couplings  needed for the models to be  cosmologically allowed without resorting to inflate walls away. In our specific $A_4$ example, the choice of $\epsilon_b$ implies constraints on the Yukawa couplings between flavons and right-handed neutrinos. 
Tables~\ref{tab:benchmarks} and~\ref{tab:benchmarks2} show some benchmark points which are in the cosmologically safe region of Fig.~\ref{fig:Main}. These points can provide a connection to other signatures, in particular the baryon asymmetry (assuming leptogenesis) and direct flavon and neutrino searches, depending on the right-handed neutrino masses. These range  from 1 GeV to the classical seesaw scale, $10^{15}$ GeV, depending on the flavor symmetry breaking scale.

The scale $v$ can be small enough for flavons to have a dynamical role at the electroweak energy scale.
Breaking scales above 1 TeV can be within the blue region, in which the peak amplitude and frequency of the gravitational waves are within existing limits or future reach of  several detectors, as shown  in Fig.~\ref{fig:GW-data}. Moreover, also part of the white region could be observed by ET and maybe LIGO, DECIGO and BBO, through the lower $f^3$ tail of the spectrum, although we have not studied in detail this possibility.

For a large range of spontaneous symmetry breaking scales, from the electroweak scale up to $10^{18}$~GeV, gravitational wave detectors  could therefore provide a characteristic signature of lepton flavor models based on non-Abelian discrete symmetries, which could not be tested in any other way. 

\acknowledgments
The work of GBG  and  EV  was  supported  by  the  U.S.  Department  of Energy (DOE) Grant No.  DE-SC0009937.
SP was supported by the European Research Council under ERC Grant NuMass (FP7-IDEAS-ERC ERC-CG 617143) and the European Union’s Horizon 2020 research and innovation programme under the Marie Sklodowska-Curie grant agreement No. 690575 (RISE InvisiblesPlus) and No. 674896 (ITN Elusives). YLZ was supported by the STFC Consolidated Grant ST/L000296/1.

\titleformat{\section}{\large\bfseries}{\appendixname~\thesection .}{0.5em}{}
\appendix
\section{Group theory of $A_4$ \label{sec:A}}

The non-Abelian group $A_4$ has three one-dimensional and one three-dimensional irreducible representations (irreps): the trivial singlet $\mathbf{1}$ and non-trivial singlets $\mathbf{1'}$, $\mathbf{1}''$ and the triplet $\mathbf{3}$. 

There are two triplet representation bases which are widely used in the literature. In the main text, we work in the Ma-Rajasekaran (MR) basis \cite{Ma:2001dn}, where the represenation matrices of $S$ and $T$ are given in Eq.~\eqref{eq:MR}. 
A triplet $a = (a_1, a_2, a_3)^T$ under actions of $S$ and $T$ transforms as
\begin{eqnarray}
S: \begin{pmatrix}
 a_1 \\ a_2 \\ a_3
\end{pmatrix} \to
\begin{pmatrix}
 a_1 \\ -a_2 \\ -a_3
\end{pmatrix} \,,\quad
T: \begin{pmatrix}
 a_1 \\ a_2 \\ a_3
\end{pmatrix} \to
\begin{pmatrix}
 a_3 \\ a_1 \\ a_2
\end{pmatrix}\,,
\end{eqnarray}
respectively. 
This basis is easier to address the vacuum structure In this basis addressing the vacuum structure is easier. However, in flavor model construction, the Altarelli-Feruglio (AF) basis \cite{Altarelli:2005yx} is more widely used. In this basis, representation matrices for $S$ and $T$ are given by 
\begin{eqnarray}
S=\frac{1}{3} \left(
\begin{array}{ccc}
 -1 & 2 & 2 \\
 2 & -1 & 2 \\
 2 & 2 & -1 \\
\end{array}
\right)\,, \quad
T=\left(
\begin{array}{ccc}
 1 & 0 & 0 \\
 0 & \omega ^2 & 0 \\
 0 & 0 & \omega  \\
\end{array}
\right) \,.
\label{eq:generators}
\end{eqnarray}
The triplet $a$ in this basis transforms  as
\begin{eqnarray}
S: \begin{pmatrix}
 a_1 \\ a_2 \\ a_3
\end{pmatrix} \to
\begin{pmatrix}
 -\frac{1}{3} a_1 + \frac{2}{3} a_2 + \frac{2}{3} a_3 \\ 
 \frac{2}{3} a_1 - \frac{1}{3} a_2 + \frac{2}{3} a_3\\ 
 \frac{2}{3} a_1 + \frac{2}{3} a_2 - \frac{1}{3} a_3
\end{pmatrix} \,,\quad
T: \begin{pmatrix}
 a_1 \\ a_2 \\ a_3
\end{pmatrix} \to
\begin{pmatrix}
 a_1 \\ \omega^2 a_2 \\ \omega a_3
\end{pmatrix}\,,
\end{eqnarray}
respectively. 
The generator $T$ is diagonal, leading to a diagonal charged lepton mass matrix, which is thus invariant under $T$ transformations. 
Therefore, the mixing matrix $U$ is directly obtained via the diagonalization of $M_\nu$. In this Appendix, Lagrangian terms mapping both bases are listed for reference. We stress that physics is independent from the choice of the basis.

The contraction of two multiplets is decomposed into 
\begin{eqnarray}
&&\mathbf{1} \times \mathbf{r} = \mathbf{r} \,,\quad
\mathbf{1}' \times \mathbf{1}'' = \mathbf{1} \,,\nonumber\\
&&\mathbf{1}' \times \mathbf{1}' = \mathbf{1}'' \,,\quad
\mathbf{1}'' \times \mathbf{1}'' = \mathbf{1}' \,,\nonumber\\
&&\mathbf{1}' \times \mathbf{3} = \mathbf{1}'' \times \mathbf{3} = \mathbf{3} \,,\nonumber\\
&&\mathbf{3} \times \mathbf{3} = \mathbf{1} + \mathbf{1}' + \mathbf{1}''+ \mathbf{3}_{S} + \mathbf{3}_{A}\,
\end{eqnarray}
where $\mathbf{r}$ can be any irrep of $A_4$, and $_S$ and $_A$ represent the symmetric and anti-symmetric contraction of the two triplets. 
The Kronecker product of two triplets $a=(a_1, a_2,a_3)^T$ and $b=(b_1, b_2, b_3)^T$ in the MR basis is given by 
\begin{eqnarray}
(ab)_\mathbf{1}\;\, &=& a_1b_1 + a_2b_2 + a_3b_3 \,,\nonumber\\
(ab)_\mathbf{1'}\, &=& a_1b_1 + \omega a_2b_2 + \omega^2 a_3b_3 \,,\nonumber\\
(ab)_\mathbf{1''} &=& a_1b_1 + \omega^2 a_2b_2 + \omega a_3b_3 \,,\nonumber\\
(ab)_{\mathbf{3}_S} &=& \frac{\sqrt{3}}{2}(a_2b_3+a_3b_2, a_3b_1+a_1b_3, a_1b_2+a_2b_1)^T \,,\nonumber\\
(ab)_{\mathbf{3}_A} &=& \frac{i}{2} (a_2b_3-a_3b_2, a_3b_1-a_1b_3, a_1b_2-a_2b_1)^T \,.
\end{eqnarray} 
In the AF basis it is given by 
\begin{eqnarray}
(ab)_\mathbf{1}\;\, &=& a_1b_1 + a_2b_3 + a_3b_2 \,,\nonumber\\
(ab)_\mathbf{1'}\, &=& a_3b_3 + a_1b_2 + a_2b_1 \,,\nonumber\\
(ab)_\mathbf{1''} &=& a_2b_2 + a_1b_3 + a_3b_1 \,,\nonumber\\
(ab)_{\mathbf{3}_S} &=& \frac{1}{2} (2a_1b_1-a_2b_3-a_3b_2, 2a_3b_3-a_1b_2-a_2b_1, 2a_2b_2-a_3b_1-a_1b_3)^T \,,\nonumber\\
(ab)_{\mathbf{3}_A} &=& \frac{1}{2} (a_2b_3-a_3b_2, a_1b_2-a_2b_1, a_3b_1-a_1b_3)^T \,.
\label{eq:CG2}
\end{eqnarray}

\section{The flavor model in two representation bases \label{sec:B}}

The general $A_4$-invariant potential for a triplet flavon, e.g., $\chi$ with the parity symmetry $\chi \to -\chi$ included, is constructed to be
\begin{eqnarray}
V_{\rm tree}(\chi)= \frac{1}{2}\mu^2_\chi (\chi \chi)_\mathbf{1} +\frac{1}{4} \left[\tilde{g}_1 \big( (\chi \chi)_\mathbf{1}\big)^2 + \tilde{g}_2 (\chi \chi)_{\mathbf{1}'} (\chi \chi)_{\mathbf{1}''} + \tilde{g}_3 \big( (\chi \chi)_{\mathbf{3}_S} (\chi \chi)_{\mathbf{3}_S} \big)_\mathbf{1} \right]\,,
\label{eq:Vvarphi}
\end{eqnarray}
Writing it explicitly in the MR basis, we obtain Eq.~\eqref{eq:self-coupling} with $g_1 = \tilde{g}_1 + \tilde{g}_2$ and $g_2 = 3(\tilde{g}_3 - \tilde{g}_2)$. 
Vacua in MR basis in Eq.~\eqref{eq:vev_chi} transform to those in the AF basis, given by 
\begin{eqnarray} 
{\bm v}_{\mathbf{1}}^{\pm}= \pm \frac{v_\chi}{\sqrt{3}} \begin{pmatrix} 1 \\ 1 \\ 1 \end{pmatrix}, ~
{\bm v}_{\mathbf{2}}^{\pm}= \pm \frac{v_\chi}{\sqrt{3}} \begin{pmatrix} 1 \\ \omega^2 \\ \omega \end{pmatrix},~
{\bm v}_{\mathbf{3}}^{\pm}= \pm \frac{v_\chi}{\sqrt{3}} \begin{pmatrix} 1 \\ \omega \\ \omega^2 \end{pmatrix}  \,.
\end{eqnarray}
Three components of $\chi$ gain masses after the spontaneous breaking of $A_4$. Although the $3 \times 3$ mass matrix varies with representation basis and vacuum we choose, three mass eigenvalues are representation- and vacuum-independent, given by $M_{\chi, 1} = \sqrt{2 g_1}\, v_\chi$, $M_{\chi, 2} = M_{\chi, 3} = \sqrt{g_2/2}\, v_\chi$. The MR basis appears to the mass basis of $\chi$, where $\chi_1$, $\chi_2$ and $\chi_3$ appear to be mass eigenstates with the diagonal mass matrix given by 
\begin{eqnarray}
M^2_\chi = 
{\rm diag} \{ M_{\chi,1}^2, M_{\chi,2}^2, M_{\chi,2}^2 \}, ~
{\rm diag} \{ M_{\chi,2}^2, M_{\chi,1}^2, M_{\chi,2}^2 \}, ~
{\rm diag} \{ M_{\chi,2}^2, M_{\chi,2}^2, M_{\chi,1}^2 \}
\end{eqnarray}
in ${\bm v}_{\bm 1}^\pm$, ${\bm v}_{\bm 2}^\pm$ and ${\bm v}_{\bm 3}^\pm$, respectively. This simple feature does not hold in the AF basis. 

The potential of $\varphi$ takes a similar form as $V_{\rm tree}(\chi)$ with the relevant coefficients replaced by $\mu_\varphi$, $\tilde{f}_1$, $\tilde{f}_2$ and $\tilde{f}_3$ with $f_1=\tilde{f}_1+\tilde{f}_2$ and $f_2=3(\tilde{f}_3-\tilde{f}_2)$. Vacua in MR basis in Eq.~\eqref{eq:vev_varphi} transform to those in the AF basis, given by
\begin{eqnarray} 
{\bm u}_{\mathbf{1}}^{\pm}= \pm v_{\varphi} \begin{pmatrix} 1 \\ 0 \\ 0 \end{pmatrix}, ~
{\bm u}_{\mathbf{2}}^{\pm}= \pm v_{\varphi} \begin{pmatrix} \frac{1}{3} \\ -\frac{2}{3} \\ -\frac{2}{3} \end{pmatrix},~
{\bm u}_{\mathbf{3}}^{\pm}= \pm v_{\varphi} \begin{pmatrix} \frac{1}{3} \\ -\frac{2}{3}\omega^2 \\ -\frac{2}{3}\omega \end{pmatrix},~
{\bm u}_{\mathbf{4}}^{\pm}= \pm v_{\varphi} \begin{pmatrix} \frac{1}{3} \\ -\frac{2}{3}\omega \\ -\frac{2}{3}\omega^2 \end{pmatrix}\,.
\end{eqnarray}
Again, the three mass eigenvalues of $\varphi$ are basis- and vacuum-independent, given by $M_{\varphi, 1} = \sqrt{2 f_1+2 f_2/3} \, v_\varphi$, $M_{\varphi, 2} = M_{\varphi, 3} = \sqrt{-f_2/3} \, v_\varphi$. But the $3\times 3$ mass matrix for $\varphi$ varies with the representation basis and vacuum we choose. The AF basis appear to be the mass basis of $\varphi$ only in the vacua ${\bm u}_{\bm 1}^\pm$, where the diagonal mass matrix is given by
\begin{eqnarray}
M^2_\varphi = 
{\rm diag} \{ M_{\varphi,1}^2, M_{\varphi,2}^2, M_{\varphi,2}^2 \}\,.
\end{eqnarray}

Examization of the flavon potential also gives rise to saddle points. The potential at these points is higher than that in the vacuum but may be lower than that at the symmetric phase. Saddle points with lowest energy density gap to
vacua are solved to be
\begin{eqnarray} \label{eq:saddle_chi}
&&
{\bm s}_{\mathbf{12}}^{\pm}= \pm r_\chi v_\chi \begin{pmatrix} 1 \\ 1 \\ 0 \end{pmatrix},~\;\;\;
{\bm s}_{\mathbf{23}}^{\pm}= \pm r_\chi v_\chi \begin{pmatrix} 0 \\ 1 \\ 1 \end{pmatrix},~\;\;\;
{\bm s}_{\mathbf{31}}^{\pm}= \pm r_\chi v_\chi \begin{pmatrix} 1 \\ 0 \\ 1 \end{pmatrix},\nonumber\\
&&
{\bm s}_{\mathbf{12}}^{\pm\prime}= \pm r_\chi v_\chi \begin{pmatrix} 1 \\ -1 \\ 0 \end{pmatrix},~
{\bm s}_{\mathbf{23}}^{\pm\prime}= \pm r_\chi v_\chi \begin{pmatrix} 0 \\ 1 \\ -1 \end{pmatrix},~
{\bm s}_{\mathbf{31}}^{\pm\prime}= \pm r_\chi v_\chi \begin{pmatrix} -1 \\ 0 \\ 1 \end{pmatrix},
\end{eqnarray}
where $r_\chi = \sqrt{M_{\chi,1}^2/(2M_{\chi,1}^2 + 2 M_{\chi,2}^2)}$. 
The energy density gap between saddle points and vacua is 
\begin{eqnarray} \label{eq:V_gap}
V_{{\rm sp},\chi} = \frac{ M_{\chi, 1}^2 M_{\chi, 2}^2 v_\chi^2}{8 (M_{\chi, 1}^2 + M_{\chi, 2}^2)} \,. 
\end{eqnarray}
It is smaller than the gap between vacua with the $A_4$-symmetric phase. The latter is calculated to be $V_{0,\chi} = M_{\chi, 1}^2 v_\chi^2/8$. 

Saddle points of $\varphi$ with lowest energy density gap are given by
\begin{eqnarray} \label{eq:saddle_varphi}
&&
{\bm r}_{\mathbf{12}}^{\pm}= \pm r_\varphi v_\varphi \begin{pmatrix} 1 \\ 1 \\ 0 \end{pmatrix},~\;\;\;
{\bm r}_{\mathbf{23}}^{\pm}= \pm r_\varphi v_\varphi \begin{pmatrix} 0 \\ 1 \\ 1 \end{pmatrix},~\;\;\;
{\bm r}_{\mathbf{31}}^{\pm}= \pm r_\varphi v_\varphi \begin{pmatrix} 1 \\ 0 \\ 1 \end{pmatrix},\nonumber\\
&&
{\bm r}_{\mathbf{12}}^{\pm\prime}= \pm r_\varphi v_\varphi \begin{pmatrix} 1 \\ -1 \\ 0 \end{pmatrix},~
{\bm r}_{\mathbf{23}}^{\pm\prime}= \pm r_\varphi v_\varphi \begin{pmatrix} 0 \\ 1 \\ -1 \end{pmatrix},~
{\bm r}_{\mathbf{31}}^{\pm\prime}= \pm r_\varphi v_\varphi \begin{pmatrix} -1 \\ 0 \\ 1 \end{pmatrix},
\end{eqnarray}
where $r_\varphi = \sqrt{(M_{\varphi, 1}^2 + 2 M_{\varphi, 2}^2)/(2 M_{\varphi, 1}^2 + M_{\varphi, 2}^2)}$. 
The energy gap between saddle points and vacua is 
\begin{eqnarray} \label{eq:V_gap}
V_{{\rm sp},\varphi} = \frac{M_{\varphi, 1}^2 M_{\varphi, 2}^2 v_\varphi^2}{8 (2 M_{\varphi, 1}^2 + M_{\varphi, 2}^2)} \,. 
\end{eqnarray}
This gap is always smaller than the gap between vacuum with the symmetric phase, which is given by $V_{0,\varphi} = M_{\varphi, 1}^2 v_\varphi^2/8$. 

Both Eqs.~\eqref{eq:saddle_chi} and \eqref{eq:saddle_varphi} are written in the MR basis. In the AR basis, these saddle points are represented as 
\begin{eqnarray} 
&&
{\bm s}_{\mathbf{12}}^{\pm}= \pm \frac{r_\chi v_\chi}{\sqrt{3}} \begin{pmatrix} 2 \\ -1 \\ -1 \end{pmatrix},~\;\;\;
{\bm s}_{\mathbf{23}}^{\pm}= \pm \frac{r_\chi v_\chi}{\sqrt{3}} \begin{pmatrix} 2 \\ -\omega^2 \\ -\omega \end{pmatrix},~\;\;\;
{\bm s}_{\mathbf{31}}^{\pm}= \pm \frac{r_\chi v_\chi}{\sqrt{3}} \begin{pmatrix} 2 \\ -\omega \\ -\omega^2 \end{pmatrix},\nonumber\\
&&
{\bm s}_{\mathbf{12}}^{\pm\prime}= \pm i r_\chi v_\chi \begin{pmatrix} 0 \\ -1 \\ 1 \end{pmatrix},~\;\;\;
{\bm s}_{\mathbf{23}}^{\pm\prime}= \pm i r_\chi v_\chi \begin{pmatrix} 0 \\ -\omega^2 \\ \omega \end{pmatrix},~\;\;\;
{\bm s}_{\mathbf{31}}^{\pm\prime}= \pm i r_\chi v_\chi \begin{pmatrix} 0 \\ -\omega \\ \omega^2 \end{pmatrix};
\end{eqnarray}
and
\begin{eqnarray} \label{eq:saddle_varphi_AF}
&&
{\bm r}_{\mathbf{12}}^{\pm}= \pm \frac{r_\varphi v_\varphi}{\sqrt{3}} \begin{pmatrix} 2 \\ -1 \\ -1 \end{pmatrix},~\;\;\;
{\bm r}_{\mathbf{23}}^{\pm}= \pm \frac{r_\varphi v_\varphi}{\sqrt{3}} \begin{pmatrix} 2 \\ -\omega^2 \\ -\omega \end{pmatrix},~\;\;\;
{\bm r}_{\mathbf{31}}^{\pm}= \pm \frac{r_\varphi v_\varphi}{\sqrt{3}} \begin{pmatrix} 2 \\ -\omega \\ -\omega^2 \end{pmatrix},\nonumber\\
&&
{\bm r}_{\mathbf{12}}^{\pm\prime}= \pm i r_\varphi v_\varphi \begin{pmatrix} 0 \\ -1 \\ 1 \end{pmatrix},~\;\;\;
{\bm r}_{\mathbf{23}}^{\pm\prime}= \pm i r_\varphi v_\varphi \begin{pmatrix} 0 \\ -\omega^2 \\ \omega \end{pmatrix},~\;\;\;
{\bm r}_{\mathbf{31}}^{\pm\prime}= \pm i r_\varphi v_\varphi \begin{pmatrix} 0 \\ -\omega \\ \omega^2 \end{pmatrix}.
\end{eqnarray}

The Lagrangian terms to generate lepton masses are given by
\begin{eqnarray}
-\mathcal{L}_{l,\nu} &\supset& y_D (\bar{L}N)_\mathbf{1} \tilde{H}  + y_N \big((\bar{N}^c N)_{\mathbf{3}_S} \phi \big)_\mathbf{1} + \frac{1}{2} u (\bar{N}^c N)_\mathbf{1} \eta \nonumber\\
&+& \frac{y_e}{\Lambda} (\bar{L} \varphi)_\mathbf{1} e_R H + \frac{y_\mu}{\Lambda} (\bar{L} \varphi)_{\mathbf{1}''} \mu_R H + \frac{y_\tau}{\Lambda} (\bar{L} \varphi)_{\mathbf{1}'} \tau_R H + \text{h.c.} \,,
\end{eqnarray}
In the AF basis, the Yukawa coupling matrix $Y_l$ after $\varphi$ gains the VEV ${\bm u}_{\bm 1}^+$ is given by 
\begin{eqnarray}
Y_l = \frac{v_\varphi}{\Lambda}\begin{pmatrix} 
y_e & 0 & 0 \\
0 & y_\mu & 0 \\
0 & 0 & y_\tau
\end{pmatrix} \,,
\end{eqnarray}
which is diagonal compared with the non-diagoanl one in Eq.~\eqref{eq:Yukawa_l} in the MR basis. 
The Dirac neutrino matrix is the same as in MR basis. Therefore, the mixing matrix is directly obtained by diagonalizing $M_N$. The latter at $\langle \chi \rangle = {\bm v}_{\bm 1}^+$ is given by
\begin{eqnarray}
M_N = \begin{pmatrix}
 u + \frac{2}{3}y_N v_\chi & -\frac{1}{3}y_N  v_\chi & -\frac{1}{3}y_N v_\chi \\
  -\frac{1}{3}y_N v_\chi & \frac{2}{3}y_N v_\chi & u - \frac{1}{3}y_N v_\chi \\
  -\frac{1}{3}y_N v_\chi & u - \frac{1}{3}y_N v_\chi & \frac{2}{3}y_N v_\chi
\end{pmatrix} \,.
\end{eqnarray}
Including the explicit breaking, $M_N$ given in the MR basis in Eq.~\eqref{eq:mass_matrices} is transformed to 
\begin{eqnarray}
M_N = \begin{pmatrix}
 u + \frac{2}{3}(y_N + \epsilon_{23}) v_\chi & -\frac{1}{3}(y_N + \epsilon_{23} - i \sqrt{3} \epsilon_{22}) v_\chi & -\frac{1}{3}(y_N + \epsilon_{23} + i \sqrt{3} \epsilon_{22}) v_\chi \\
  -\frac{1}{3}(y_N + \epsilon_{23} - i \sqrt{3} \epsilon_{22}) v_\chi & \frac{1}{3}(2y_N + 2\epsilon_{23} - i\sqrt{3} \epsilon_{22}) v_\chi & u - \frac{1}{3} (y_N + \epsilon_{23}) v_\chi \\
  -\frac{1}{3}(y_N + \epsilon_{23} + i \sqrt{3} \epsilon_{22}) v_\chi & u - \frac{1}{3} (y_N + \epsilon_{23}) v_\chi & \frac{1}{3}(2y_N + 2\epsilon_{23} + i\sqrt{3} \epsilon_{22}) v_\chi
\end{pmatrix} \,. \nonumber\\
\end{eqnarray}
in the AF basis. This matrix satisfies $(M_N)_{e\mu} = (M_N)_{e\tau}^*$, $(M_N)_{\mu\mu} = (M_N)_{\tau\tau}^*$ and $(M_N)_{\mu\tau} = (M_N)_{\mu\tau}^*$ \cite{Babu:2002dz}.

\bibliography{gwflavons}
\bibliographystyle{bibi}

\end{document}